\documentclass[11pt]{article}
\pdfoutput=1
\usepackage{jcapmod}
\usepackage{booktabs}
\usepackage{mathtools}
\usepackage{setspace} 
\usepackage[english]{babel}
\usepackage{amsmath,amssymb,amsbsy,amstext, amsthm,xcolor}
\usepackage{graphicx}
\usepackage{amsfonts}
\usepackage{amssymb}
\usepackage{verbatim}
\usepackage{float}
\usepackage[utf8]{inputenc}
\usepackage{caption}
\usepackage{subcaption}
\usepackage{enumitem}
\RequirePackage{color}

\usepackage{colortbl}
\definecolor{green2}{cmyk}{0, 1, 0.5, 0}
\definecolor{lightgreen}{cmyk}{0.2, 0, 0.2, 0.2}
\definecolor{lightgray}{cmyk}{0.1,0.2,0,0.1}
\definecolor{lightgray2}{cmyk}{0.4,0.4,0,0.8}
\definecolor{black}{cmyk}{1.0,1.0,1.0,1.0}

\allowdisplaybreaks[1]

\usepackage{colortbl}

\definecolor{lightgreen}{cmyk}{0.2, 0, 0.2, 0.2}
\definecolor{lightgray}{cmyk}{0.1,0.2,0,0.1}
\definecolor{lightgray2}{cmyk}{0.1,0.1,0,0.1}

\makeatletter
\newlength{\apb@width}
\newcommand{\autoparbox}[2][c]{\settowidth{\apb@width}{#2}\parbox[#1]{\apb@width}{#2}}

\makeatother

\numberwithin{equation}{section}
\def\beq{\begin{equation}}
\def\eeq{\end{equation}}

\def\bea{\begin{eqnarray}}
\def\eea{\end{eqnarray}}
\def\eg{{\it e.g.~}}

\def\ie{{\it i.e.~}}
\def\d{{\rm d}}

\def\d{{\rm d}}

\def\nn{\nonumber}

\def\sgm{\sigma}

\def\del{\partial}
\def\Mp{M_{\rm pl}}

\def\fr{\frac}

\def\0{{\boldsymbol 0}}

\def\fr{\frac}

\begin{document}

\begin{titlepage}

\setcounter{page}{1} \baselineskip=15.5pt \thispagestyle{empty}

\bigskip\

\vspace{1cm}
\begin{center}

{\fontsize{20}{28}\selectfont  \sffamily \bfseries {Primordial black holes as dark matter and \smallskip

 gravitational waves from bumpy axion inflation
}}

\end{center}

\vspace{0.2cm}

\begin{center}
{\fontsize{13}{30}\selectfont Ogan \"Ozsoy$^{\star}$ and Zygmunt Lalak$^{\star}$
}
\end{center}

\begin{center}

\vskip 8pt
\textsl{
$\star$ Institute of Theoretical Physics, Faculty of Physics, University of Warsaw, ul. Pasteura 5, Warsaw, Poland}
\vskip 7pt

\end{center}

\vspace{1.2cm}
\hrule \vspace{0.3cm}
\noindent 
We consider a mechanism for producing a significant population of primordial black holes (PBHs) and an observable stochastic gravitational wave background (SGWB) within string theory inspired models of inflation. In this framework where inflaton is identified as a non-compact axion-like field, sub-leading non-perturbative effects can superimpose steep cliffs connected by smooth plateaus onto the underlying axion potential. In the presence of coupling to Abelian gauge fields, the motion of axion on the cliff-like region(s) of its potential triggers a localized production of one helicity state of gauge fields due to the temporary fast-roll of axion around such a feature. In this setup, primordial fluctuations sourced by vector fields exhibit a localized peak in momentum space corresponding to modes that exit the horizon when the axion velocity is maximal. As an application of this general mechanism, we present an example of axion inflation which both matches Planck observations at CMB scales and generates a population of light PBHs ($M_{\rm PBH} \simeq 10^{-13} M_{\odot}$) that can account for all dark matter. In this scenario, the enhanced scalar fluctuations that leads to PBHs also generate an observable SGWB of induced origin at LISA scales. The amplitude and shape of the resulting GW signal inherits specific properties (such as non-Gaussianity and its shape) of its scalar sources that may allow us to distinguish this mechanism from other inflationary scenarios and astrophysical backgrounds. This GW signal together with an observation of PBH distribution at the corresponding scales can thus provide a window to the inflationary dynamics on scales much smaller than those probed by Cosmic Microwave Background (CMB) and Large Scale Structure (LSS) Measurements. 
\vskip 10pt
\hrule

\vspace{0.6cm}
 \end{titlepage}

 \tableofcontents
 
\newpage

\section{Introduction}

Observations on CMB and LSS strongly support the inflationary paradigm in the early universe \cite{Akrami:2018odb,Dawson:2015wdb,Cuesta:2015mqa}. While these observations\footnote{Spectral distortion experiments can further push this range up to $k \lesssim 10^{4}\, {\rm Mpc}^{-1}$ \cite{Fixsen:1996nj,Kogut:2011xw,Andre:2013nfa}.} allow us to probe the inflationary dynamics through the largest cosmological scales, $10^{-4}\, {\rm Mpc}^{-1} \lesssim k \lesssim 10^{-1}\, {\rm Mpc}^{-1}$ corresponding to $60-50$ e-folds before the end of inflation, we do not have direct access to inflationary dynamics on small scales except\footnote{Bounds on the abundance of ultracompact minihalos may also lead to additional constraints \cite{Bringmann:2011ut,Emami:2017fiy}.} for bounds on PBHs which arise if the scalar fluctuations have a sufficiently large amplitude at small scales \cite{Hawking:1971ei,Carr:1974nx}. Constraints from various physical processes on PBH abundance continue to improve but leave viable windows especially when astrophysical uncertainties are taken into account (See \cite{Carr:2016drx,Garcia-Bellido:2017fdg,Sasaki:2018dmp,Carr:2020xqk,Green:2020jor} for recent reviews).

Excitingly, PBHs could account for a significant fraction or totality of mysterious dark matter (DM) density that dominates cosmic structures in the present day universe. In particular, recent observations of gravitational waves (GWs) \cite{Guzzetti:2016mkm,Caprini:2018mtu} by black hole mergers as well as the absence of astrophysical and collider signatures for well-motivated particle DM candidates rekindled this idea which is observationally viable for PBHs within the mass range of  $10^{-16}\,\lesssim M_{\rm PBH} \, [M_{\odot}]\lesssim 10^{-12}\,  $ (corresponding to $5 \times 10^{11}\,  \lesssim k \, [{\rm Mpc}^{-1}] \lesssim 5 \times 10^{14}\, $) as discussed recently in \cite{Kawasaki:2016pql,Niikura:2017zjd,Katz:2018zrn,Montero-Camacho:2019jte} (See also \cite{Laha:2019ssq,Dasgupta:2019cae,Laha:2020ivk}). The most compelling modern process of PBH formation is related to the enhancement of super-horizon curvature perturbations \cite{Leach:2000yw,Leach:2001zf,Ozsoy:2019lyy} that originated as quantum fluctuations during inflation: Upon horizon reentry, these large fluctuations collapse during the radiation dominated era to form black holes with masses of the order of mass contained within the horizon at horizon re-crossing.

Many recent works have explored various primordial mechanisms on how such an enhancement could be achieved, including: the presence of features in the scalar potential (an inflection point or a sudden change in its slope) in single field inflation \cite{Starobinsky:1992ts,Ivanov:1994pa,Garcia-Bellido:2017mdw,Motohashi:2017kbs,Ballesteros:2017fsr,Cicoli:2018asa,Ozsoy:2018flq,Dalianis:2018frf,Motohashi:2019rhu,Mishra:2019pzq,Ballesteros:2019hus,Ballesteros:2020qam}, through the instability of a scalar fields during inflaton \cite{Garc_a_Bellido_1996,Clesse:2015wea,Garcia:2020mwi} (See \cite{Braglia:2020eai,Palma:2020ejf,Fumagalli:2020adf} for other interesting multi-field scenarios), from gauge field sources amplified by a rolling axion \cite{Linde:2012bt,Bugaev:2013fya,Erfani:2015rqv, Garc_a_Bellido_2016,Domcke:2017fix,Almeida:2020kaq}, multiple stages of inflation with a short break (or temporary halt) of inflation \cite{Polarski:1992dq,Polarski:1994rz,Saito:2008em,Kannike:2017bxn,Ozsoy:2018flq}, from small speed of sound \cite{Ozsoy:2018flq,Ballesteros:2018wlw} and the resonance in the speed of sound of curvature perturbations during inflation \cite{Cai:2018tuh,Chen:2020uhe}.

A common feature of all early universe scenarios that leads to PBH formation is the inevitable production of a stochastic GW background (SGWB) due to gravitational coupling of enhanced scalar fluctuations with tensor modes at second order in perturbation theory \cite{Mollerach:2003nq,Ananda:2006af,Osano:2006ew,Baumann:2007zm,Kohri:2018awv}: although scalar-tensor interaction is of gravitational strength (\ie Planck suppressed), the enhancement of scalar perturbations required to produce PBH can induce a significant amount of GWs as the scalar modes re-enter the horizon in the radiation dominated universe\footnote{For a partial list of models that studies induced SGWB produced from scalar fluctuations enhanced during inflation, see \cite{Ballesteros:2020qam,Braglia:2020eai,Lin:2020goi,Fu:2019vqc,Drees:2019xpp,Cai:2019jah,Cai:2019bmk,Espinosa_2018}.}. Interestingly, this signal contains crucial information about the properties of its sources, namely the amplitude and statistics of scalar perturbations: for an equal amount PBH population of certain mass, a smaller SGWB is obtained for non-Gaussian scalar perturbations compared to the Gaussian primordial curvature perturbation modes \cite{Garc_a_Bellido_2017}. In this sense, the determination of present PBH mass distribution together with its associated GW signal contains key information on the statistics of these modes complementary to the CMB probes and can help us to distinguish between different models on the origin of these fluctuations. Considering the sensitivity next generation spatial based experiments such as LISA \cite{Caprini:2015zlo,Bartolo:2016ami} will reach, simultaneous observation of these signals provide us an opportunity to access inflationary dynamics on scales much smaller than those currently probed with CMB and LSS experiments.

In light of this information, our main objective in this work is to identify a string-inspired mechanism that can give rise to scalar (strongly non-Gaussian) scalar and tensor fluctuations during inflation, capable of generating significant population of PBHs together with a SGWB that typically involves multiple components including the induced GWs at second-order in perturbation theory. For concreteness, we consider a string-inspired model of axion inflation, \eg axion monodromy with drifting oscillations \cite{McAllister:2008hb,McAllister:2014mpa,Flauger:2014ana}, where the discrete shift symmetry of the axion is broken both by a non-periodic monomial term plus a drift factor multiplying axion modulations \cite{Kobayashi:2015aaa,BLZ,Kallosh:2014vja}:
\beq\label{Vb}
V(\phi) = \fr{1}{2}m^2 \phi^2 + \Lambda^4~ \fr{\phi}{f}~\sin\left(\fr{\phi}{f}\right).
\eeq
For sizeable modulations $\Lambda^{4} \lesssim m^2 f^2$ (which we refer as ``bumpy regime'' in what follows), the last term in \eqref{Vb} introduces plateau like regions connected by steep cliffs on to the underlying potential (See Figure \ref{fig:V}). Focusing on this regime, earlier works have shown that the modified dynamics in axion inflation can lead to interesting phenomenology at CMB scales including: running of the spectral index \cite{Kobayashi:2010pz}, prolonged duration of inflation with intermediate (super-Planckian) field ranges and relatively small tensor-to-scalar ratio \cite{Parameswaran:2016qqq}. In \cite{Ozsoy:2018flq}, it was also shown that the presence of sizeable modulations may also introduce a feature (namely a shallow local minimum followed by an inflection point) in the scalar potential at small field values, leading to a pronounced peak in the scalar perturbations required for PBH formation at sub-CMB scales. In this model, as in all single field models of inflation, the required enhancement of scalar perturbations (hence the PBH abundance) is highly sensitive on model parameters that control the depth of the local minimum of the scalar potential \cite{Ballesteros:2017fsr,Hertzberg_2018}. On the other hand, as in the model we study here, tuning the parameters of the scalar potential does not always guarantee the conditions to generate a pronounced peak in the scalar perturbations. In order to mitigate these shortcomings associated with PBH formation in single field models of inflation, a reasonable price one can pay is to consider the presence of additional sectors that exhibit couplings to inflaton. In this context, inflation driven by axion-like fields appear as a natural candidates because due to their approximate shift symmetry they are expected to couple to gauge fields through a dimension five operator\footnote{Shift symmetric scalars can also couple to fermion current through dimension five operators. See \cite{Adshead:2015kza,Adshead:2018oaa,Domcke:2018eki,Adshead:2019aac} for theoretical and phenomenological implications of such coupling during axion inflation.}:
\beq\label{LINT}
\fr{\Delta\mathcal{L}_{\rm int}}{\sqrt{-g}} = -\fr{\alpha_c}{4f} \phi F\tilde{F},
\eeq
where $F$ is field-strength tensor, $\tilde{F}$ is its dual and $\alpha_{\rm c}/f$ controls the size of the coupling with axion $\phi$, $f$ being the axion decay constant. The coupling \eqref{LINT}  breaks parity in the gauge field sector and leads to an amplification of one helicity state of gauge field fluctuations $\propto {\rm e}^{\dot{\phi}/Hf}$ giving rise to inflationary dynamics with a rich set of phenomenological consequences\footnote{A partial list includes inflation on a steep potential \cite{Anber:2009ua}, magnetogenesis during inflation \cite{Anber:2006xt,Caprini:2014mja}, large scalar \cite{Barnaby:2010vf,Barnaby:2011vw} and tensor non-Gaussianity at CMB scales \cite{Cook:2013xea,Agrawal:2017awz,Agrawal:2018mrg,Ozsoy:2020ccy}, parity violation in the CMB \cite{Sorbo:2011rz,Shiraishi:2013kxa,Adshead:2013qp} and interferometer scales \cite{Thorne:2017jft} and efficient preheating \cite{Adshead:2015pva,Cuissa:2018oiw} that contributes to the effective number of relativistic degrees of freedom $\Delta N_{\rm eff}$ \cite{Adshead:2019igv,Adshead:2019lbr}. Identifying axion as a spectator may also lead to inflationary scenarios with observable non-Gaussian GWs at CMB scales from secondary gauge field sources \cite{Namba:2015gja,Ozsoy:2020ccy}.} including the production of primordial black holes \cite{Linde:2012bt,Bugaev:2013fya,Erfani:2015rqv, Garc_a_Bellido_2016,Domcke:2017fix,Almeida:2020kaq}. In most of the previous literature that utilize the coupling \eqref{LINT} during inflation, axion potentials that give rise to smooth and monotonically increasing effective coupling $\xi \propto \dot{\phi}/Hf$ was considered to enhance scalar and tensor fluctuations at sub-CMB scales through gauge field sources.  Recently, an exception to this appeared in \cite{Cheng:2016qzb,Cheng:2018yyr} where localized enhancement\footnote{Similarly, with an aim to generate visible GWs at interferometer scales, tensor fluctuations that exhibit a localized blue tilt can be obtained through transient non-attractor phases in scalar-tensor theories of single field inflation \cite{Mylova:2018yap,Ozsoy:2019slf}.} of scalar and tensor perturbations (from gauge field sources) is studied numerically in a model of axion inflation that utilizes sizeable \emph{constant} modulations in the scalar potential. In the present work however, we consider inflation with axion-like field where the potential exhibit drifting modulations (See eq. \eqref{Vb}) which allow us to initiate an accurate semi-analytic study of enhancement in primordial fluctuations in the presence of the coupling in eq. \eqref{LINT} (See \eg Appendix \ref{AppA} and \ref{AppC}).   

The principle mechanism that give rise to an enhancement of primordial fluctuations is as follows: within each step like feature, the velocity of $\phi$, \ie $\dot{\phi}/Hf$ is very small in the plateau regions of the potential \eqref{Vb} whereas it transiently peaks in the cliff-like regions connecting to plateaus. The transient increase in $\dot{\phi}/Hf$ around such a feature triggers a localized production of gauge field fluctuations which in turn sources scalar and tensor fluctuations through inverse decay processes: $\delta A + \delta A \to \delta \phi$ and $\delta A + \delta A \to h$. In contrast to the continuous particle production scenarios, localized nature of particle production we consider in this work has the advantage of inducing negligible back-reaction (See Appendix \ref{AppE}) on the motion of inflation. This stems from the fact that for monotonic inflaton potentials the coupling $|\dot{\phi}|/Hf$ that controls the efficiency of particle production is increasing continuously during inflation and once it reaches a critical value system enters in a strong back-reaction\footnote{See \eg \cite{Cheng:2015oqa} and \cite{Domcke:2020zez} for a recent study on interesting features associated with the strong backreaction regime in axion inflation.} regime.  

In this paper, we focus on a representative parameter space in the bumpy regime to analyze in detail the CMB and sub-CMB phenomenology that arise in axion inflation when the coupling \eqref{LINT} between axion and gauge fields is present. In light of current uncertainties on PBH limits, we will focus our attention on the scales relevant for the forthcoming LISA mission\footnote{The possibility to test PBH dark matter with LISA is first discussed in \cite{Garc_a_Bellido_2017}. See also \cite{Bartolo:2018evs,Bartolo:2018rku} for a general discussion including tensor non-Gaussianities in this context.} to study the sub-CMB phenomenology in this inflationary scenario. 

The organization of this paper is as follows: In Section \ref{Sec2}, we introduce the bumpy regime in axion inflation and study gauge field production as the axion traverses steep the cliff(s) in its wiggly potential. In Section \ref{SF}, we review the dynamics of primordial fluctuations in the presence of gauge field sources. In Section \ref{Sec4}, focusing on an explicit numerical example of background evolution, we study the CMB and sub-CMB phenomenology in the bumpy axion inflation, with an emphasis on production of PBHs and SGWB at LISA scales. In Section \ref{Sec5} we present our conclusions.

\section{The model}\label{Sec2}
Following the discussion in the introduction, we consider a model of axion-like field $\phi$ with canonical kinetic term and an abelian gauge field sector where these sectors talk to each other through Chern-Simons type coupling and both sectors minimally coupled to the Einstein gravity. The action for the system is given by:
\beq\label{ML}
\fr{\mathcal{L}}{\sqrt{-g}}=\fr{\Mp^2}{2}R-\frac{1}{2}\partial_{\mu} \phi\partial^{\mu} \phi-V(\phi)-\fr{1}{4}F_{\mu\nu}F^{\mu\nu}-\fr{\alpha_{\rm c}}{4f} \phi F_{\mu\nu}\tilde{F}^{\mu\nu}\eeq
where $V(\phi)$ is the scalar potential for $\phi$ introduced in eq. \eqref{Vb} and $f$ is its decay constant and $\alpha_{\rm c}$ is dimensionless constant that determines the strength of the coupling to the gauge fields. Here, the gauge field strength tensor and its dual are defined by $F_{\mu\nu} = \del_\mu A_\nu - \del_\nu A_\mu$ and $\tilde{F}^{\mu\nu}\equiv \eta^{\mu\nu\rho\sigma} F_{\rho\sigma}/(2\sqrt{-g})$ where alternating symbol $\eta^{\mu\nu\rho\sigma}$ is $1$ for even permutation of its indices, $-1$ for odd permutations, and zero otherwise.
\subsection{Bumpy Regime of Axion Inflation}

\begin{figure}[t!]
\begin{center}
\includegraphics[scale=0.63]{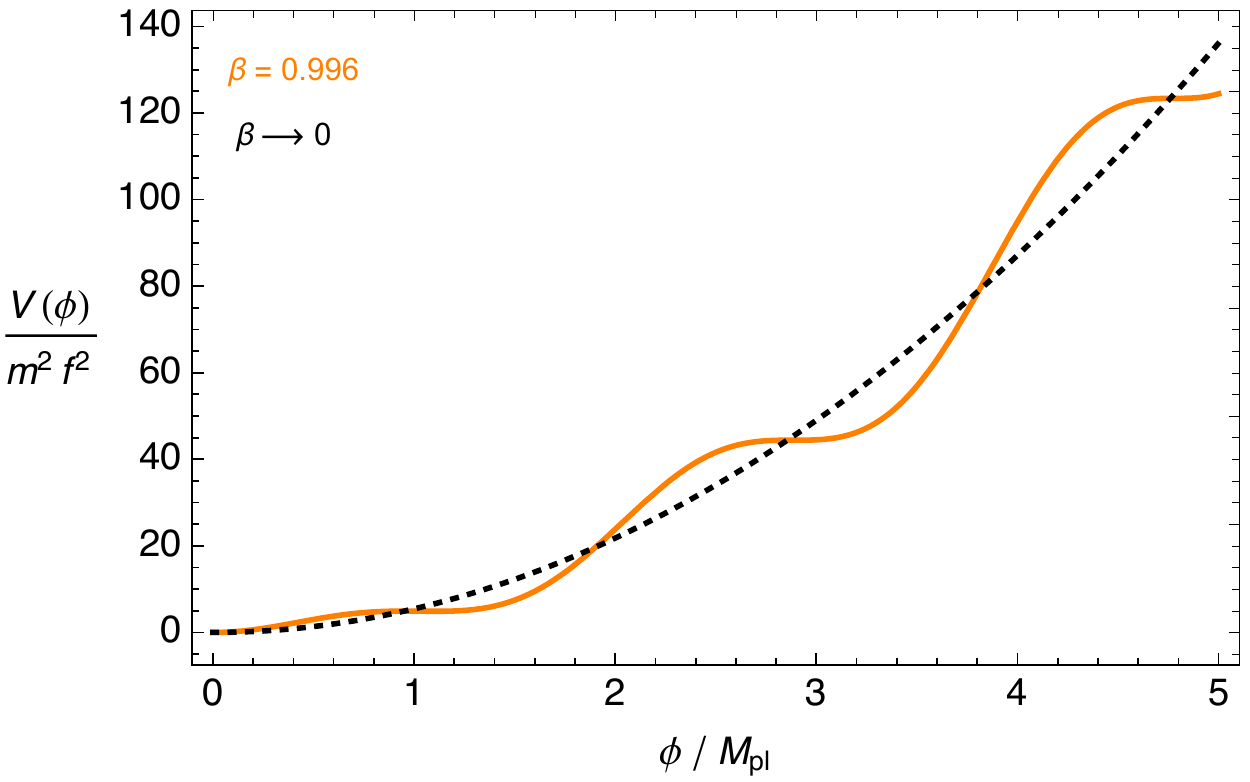}\includegraphics[scale=0.63]{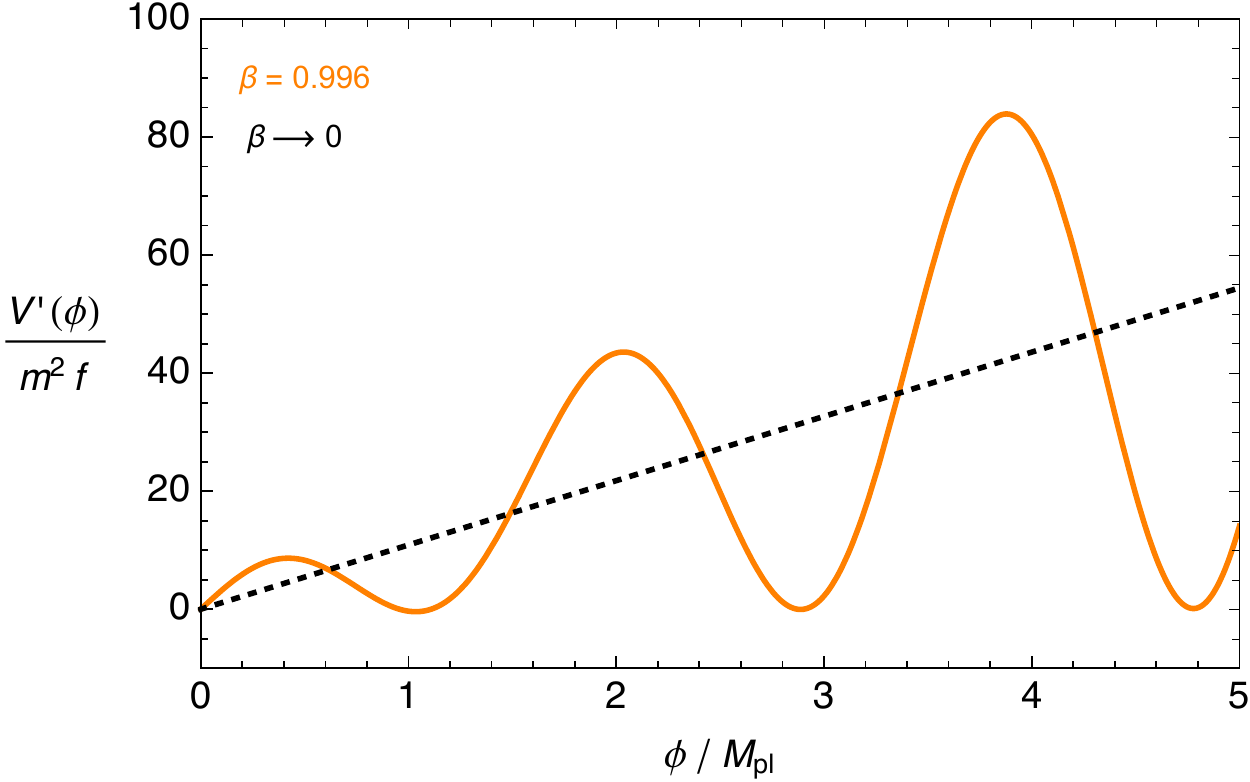}
\end{center}
\caption{Potential $V(\phi)$ (left) in \eqref{Vb} and its derivative $V'(\phi)$ (right) for parameters $\beta 
\equiv \Lambda^4/m^2f^2 = 0.996$ and $\Mp/f = 3.3$. For comparison we also plot the potential in the $\beta \to 0$ limit, \ie for smooth quadratic potential $V(\phi)\propto \phi^2$ (black, dotted).\label{fig:V}}
\end{figure}

In this section our aim is to describe the background evolution of inflaton when its scalar potential exhibit sizeable axion modulations in addition to the monomial term (See Figure \ref{fig:V}). In particular, our main focus will be the axion field profile around the step like features which is required to set the stage for gauge field production we study in the next section.

In effective descriptions of axion inflation based on string theory compactifications, the continous shift symmetry of the axion can be spontaneously broken by background vevs (\eg fluxes) and/or non-perturbative effects (\eg string instantons), leading to large field inflation models with monomial and/or cosine (``natural inflation'' \cite{Freese:1990rb}) potentials. If the non-perturbative corrections are sufficiently large they can introduce sizeable modulations into the underlying potential. The size of these effects will depend on the details of the microscopic data, in particular on the vev's of fluxes and other moduli that are already stabilised. Therefore, they can induce small oscillations \cite{Flauger:2009ab} on to the potential or dominant\footnote{Arguments \cite{Hebecker:2015zss} based on Weak Gravity Conjecture (WGC) \cite{ArkaniHamed:2006dz} can be used constraint the size of the modulations in axion monodromy potential. For the model we are considering, these theoretical considerations imply $\beta < \Mp^2 /f^2$. For sub-Planckian axion decay constants, this upper bound is automatically satisfied considering the mild bumpy regime $\beta < 1$ we are operating in this work.} enough to introduce new local minima and maxima that may halt inflation \cite{Banks:2003sx}. In this work, we will consider an intermediate situation, where sizeable but sub-dominant non-perturbative corrections introduces step-like features in the potential including steep cliffs and gentle plateaus.

The homogeneous dynamics of the axion depends on the size of the non-perturbative corrections compared to the mass term in the potential \eqref{Vb}, in particular on the ratio $\beta= \Lambda^4 / (m^2 f^2)$. In the regime we are interested in, non-perturbative effects in the scalar potential are sufficiently large but appear as a sub-dominant piece corresponding to $\beta \lesssim 1$, without assuming $\beta \ll 1$. In this regime, we illustrate the global shape of the potential and its slope in Figure \ref{fig:V} for a representative choice of parameters. We observe that the presence of sizeable non-perturbative corrections introduce plateau-like regions connected by steep cliffs onto the underlying axion potential. Notice also that at large field values, the slope of the potential $V'(\phi)$ exhibits deep wells/high barriers, indicating regions in the potential that have smaller/larger slopes compared to the standard quadratic potential. An initially displaced $\phi$ would roll down in its wiggly potential where it transiently speeds up in the cliff like regions before slowing down in smooth plateaus and eventually settling on its global minimum at $\phi=0$ \cite{Parameswaran:2016qqq}. Depending on the initial conditions and model parameters, $\phi$ might probe multiple cliffs of the scalar potential. In this work, we will work with a parameter space in which plateau like regions are flat enough to allow the axion complete its entire $60$ e-folds of evolution within one such feature (See \eg Section \ref{Sec4p2}). In the following, we will briefly describe the homogeneous evolution of $\phi$ during inflation while it traverses one such step like region in its scalar potential. For more details regarding background evolution including the approximations we undertake, see Appendix \ref{AppA}.

{\bf Background evolution through the bumps:} Assuming potential energy $V(\phi)$ in \eqref{Vb} dominates the energy budget of the universe during inflation, \ie $3 H^2 \Mp^2 \simeq V(\phi)$, one can derive simple analytic expressions that describe the dynamics of the axion-like field $\phi$. Without making a slow-roll approximation, within a step like region in its potential including two plateau like regions separated by a cliff, we obtain 
\beq\label{dphi}
 \fr{\dot{\phi}}{2Hf} = - \fr{\delta}{1+\delta^2 (N-N_*)^2},
\eeq
where $\delta \equiv \alpha(1+\beta)(m/\sqrt{6} H)$ is constant dimensionless parameter assuming an approximately constant Hubble rate $H$, $N$ denotes e-folds with $N_*$ representing the e-folding number when the velocity of the axion field in \eqref{dphi} reaches its peak value. We observe from \eqref{dphi} that axion has a non-negligible velocity on for a limited amount of e-folds given by $\Delta N =N - N_* \sim \delta ^{-1}$. During these times axion is rolling over the cliff like region in its potential while its velocity becomes increasingly small at the plateaus, \ie when $|\Delta N| \gg 1 $. 

In the presence of the coupling \eqref{LINT} to the gauge fields, the kinetic energy of the axion acts as a source for the gauge field fluctuations and amplifies its vacuum fluctuations \cite{Anber:2009ua}. The efficiency of this process is controlled by the dimensionless ``effective coupling'' $\xi \equiv - \alpha_c\dot{\phi}/(2Hf)$ which must be larger than unity in order to lead to significant particle production in the gauge field sector. In the next subsection, we will focus on the amplification of gauge field fluctuations as the axion-like field traverse the step like parts of its potential where it exhibits the velocity profile in eq. \eqref{dphi}. 
\subsection{Gauge Field Production}\label{Sec2p2}
The equation of motion for the gauge field can be obtained by varying the action in \eqref{ML} in Coulomb gauge,
\beq\label{EMG}
A_i'' - \vec{\nabla}^2 A_i - \fr{ \alpha_{\rm c} a(\tau)\dot{\phi}}{f} ~\epsilon_{ijk}~ \del_j A_k = 0.
\eeq
We decompose the gauge field $A_i$ in terms of the annihilation and creation operators in the usual way,
\beq\label{DGF}
A_i(\tau, \vec{x}) = \int \fr{\d^3 k}{(2\pi)^{3/2}} ~ e^{i\vec{k}.\vec{x}}  \sum_{\lambda = \pm} \epsilon^{\lambda}_i(\vec{k}) \left[
A_\lambda(\tau,\vec{k})\hat{a}_\lambda(\vec{k}) + A^{*}_\lambda(\tau,-\vec{k})\hat{a}^{\dagger}_\lambda(-\vec{k})  \right],   
\eeq
where the helicity vectors obey $k_i \epsilon^{\pm}_i = 0$, $\epsilon_{ijk}~ k_j ~\epsilon^{\pm}_k = \mp i k \epsilon^{\pm}_i$, $\epsilon^{\pm}_i\epsilon^{\pm}_i = 0$, $\epsilon^{\pm}_i \epsilon^{\mp}_i =1$ and $(\epsilon^{\lambda}_i(\vec{k}))^{*} = \epsilon^{\lambda}_i(-\vec{k}) = \epsilon^{-\lambda}_i(\vec{k})$ and the annihilation/creation operators satisfy $\left[\hat{a}_\lambda(\vec{k}),\hat{a}^\dagger_{\lambda'}(\vec{k}')\right] = \delta_{\lambda\lambda'} \delta^{(3)}(\vec{k}-\vec{k}')$. 

Using the decomposition \eqref{DGF} in \eqref{EMG}, the mode functions $A_\lambda$ can be shown to obey 
\beq\label{MEA}
A_{\pm}''(x) + \left(1\pm \fr{2\xi(x)}{x}\right)A_\pm(x) =0,
\eeq
where we defined dimensionless variable $-k\tau = x$. Realize that with our conventions ($\dot{\phi} < 0$ or $\xi >0$), time dependent mass term in \eqref{MEA} can trigger tachyonic instability only for the negative helicity state $A_-$ for modes satisfying $-k\tau < 2\xi$. 

In the present work, we need to solve eq. \eqref{MEA} when $\xi$ exhibit the profile given in eq. \eqref{xiapp}. Using a semi-analytic procedure we explain in Appendix \ref{AppA}, an explicit expression for the late time dependence of $A_{-}(\tau,k)$ can be obtained \cite{Ozsoy:2020ccy}:

\beq\label{Am}
A_{-}(\tau, k) \simeq\left[\frac{-\tau}{8 k \xi(\tau)}\right]^{1 / 4} \tilde{A}(\tau, k), \quad A_{-}^{\prime}(\tau, k) \simeq\left[\frac{k \xi(\tau)}{-2 \tau}\right]^{1 / 4} \tilde{A}(\tau, k)
\eeq
where we defined 
\beq\label{tA}
 \tilde{A}(\tau, k) = N(\xi_*,x_*,\delta)\exp \left[- \fr{2\sqrt{2\xi_*}~ (-k\tau)^{1/2} }{\delta |\ln(\tau/\tau_*)| }\right], \,\,\,\,\,\,\,\,\,\quad\quad\quad\quad \tau/\tau_* < 1.
\eeq
In eq. \eqref{tA}, the normalization factor (real and positive) $N(\xi_*,x_*,\delta)$ parametrizes the scale dependence and the sensitivity of the mode functions on the background dynamics  through $x_* =- k\tau_* = k/k_*$, $\xi_*$ and $\delta$ with $\tau_*$ denoting the conformal time when $\xi$ reaches its peak value $\xi_*$ while axion rolls through the cliffs. 

Before we finalize this section, in analogy with Standard Model notation, we define ``Electric'' and ``Magnetic" fields in terms of the auxiliary potential $A_i$: $E_i = -{a^{-2}}~ A_i' , ~~~ B_i = {a^{-2}}~\epsilon_{ijk}~\del_j A_k$. The
Fourier transforms of these sources are given by (See Section \ref{SF}): 
\begin{align}\label{EBF}
\nn \hat{E}_i(\tau,\vec{k}) &= -(H\tau)^2 \sqrt{\fr{k}{2}} ~\epsilon_{i}^{-}(\vec{k}) \left(\fr{2\xi(\tau)}{-k\tau}\right)^{1/4} \tilde{A}(\tau,k)\left[\hat{a}_{-}(\vec{k})+~\hat{a}^{\dagger}_{-}(-\vec{k})\right],\\
\hat{B}_i(\tau,\vec{k}) &= -(H\tau)^2 \sqrt{\fr{k}{2}}~ \epsilon_{i}^{-}(\vec{k}) \left(\fr{-k \tau}{2\xi(\tau)}\right)^{1/4} \tilde{A}(\tau,k)\left[\hat{a}_{-}(\vec{k})+~\hat{a}^{\dagger}_{-}(-\vec{k})\right],
\end{align}
where we have used \eqref{DGF}, \eqref{Am} and the definitions of electric and magnetic fields above.
\section{Primordial fluctuations sourced by vector fields during inflation}\label{SF}
The Lagrangian \eqref{ML} contains one scalar and two tensor modes as dynamical variables. To linear order in perturbations, we decompose these fluctuations as
\begin{align}
\label{pdc}\hat{\phi}(\tau, \vec{x}) &=\phi(\tau)+\int \frac{\d^{3} k}{(2 \pi)^{3 / 2}}~ \mathrm{e}^{i \vec{k} \cdot \vec{x}}~ \frac{\hat{Q}_{\phi}(k, \tau)}{a(\tau)},\\
\label{tdc}\hat{h}_{ij}(\tau, \vec{x}) &= \fr{2}{\Mp}\int \fr{\d^3 k}{(2\pi)^{3/2}}~ \mathrm{e}^{i\vec{k}.\vec{x}} \sum_{\lambda=\pm} \Pi^{*}_{ij,\lambda} \fr{\hat{Q}_{\lambda}(\tau,\vec{k})}{a(\tau)},
\end{align}
where $h_{ij}$ is the transverse, $\partial_i h_{ij} = 0$ and traceless, $h_{ii} =0$ metric perturbation and the polarization operators are defined as $\Pi^{*}_{ij,\pm} = \epsilon^{\pm}_i(\vec{k}) \epsilon^{\pm}_j(\vec{k}), ~\Pi_{ij,\pm} = \epsilon^{\mp}_i(\vec{k}) \epsilon^{\mp}_j(\vec{k})$, satisfying $\Pi^{*}_{ij,\lambda}\Pi_{ij,\lambda'} = \delta_{\lambda\lambda'}$.  In the spatially flat gauge, the metric is given by
\beq\label{MA}
\mathrm{d} s^{2}=a^{2}(\tau)\left[-N^{2} \mathrm{d} \tau^{2}+ (\delta_{i j}+h_{ij})\left(\mathrm{d} x^{i}+N^{i} \mathrm{d} 
\tau\right)\left(\mathrm{d} x^{j}+N^{j} \mathrm{d} \tau\right)\right],
\eeq 
where $N = 1 +\delta N$ and $N^{i}$ are non-dynamical lapse and shift function respectively. Plugging the metric in eq. \eqref{MA} into the Lagrangian \eqref{ML}, one can solve for the lapse and shift in terms of the dynamical scalar mode (See \eg \cite{Barnaby:2011vw, Ozsoy:2017blg}). In this way, the action for physical scalar fluctuation $Q_\phi$ can be shown to assume the following form,
\beq\label{SQp}
S\left[\hat{Q}_{\phi}\right]=\frac{1}{2} \int \d \tau \d^{3} k\left\{\hat{Q}'_{\phi}\hat{Q}_{\phi}'-\left[k^{2}+m^2_{\mathrm{eff}} (\tau)\right] \hat{Q}^2_{\phi} + 2\hat{Q}_\phi ~ \hat{J}_\phi (\tau, \vec{k})\right\},
\eeq
where $m^2_{\mathrm{eff}} $ a time dependent mass that we will further elaborate on in the following section and  $\hat{J}_\phi$ is the source induced due to the couplings to the gauge fields: 
\beq
\hat{J}_\phi (\tau, \vec{k}) \equiv \frac{\alpha_{\mathrm{c}} a(\tau)^{3}}{f} \int \frac{\mathrm{d}^{3} x}{(2 \pi)^{3 / 2}}~\mathrm{e}^{-i \vec{k} \cdot \vec{x}}~ \hat{E}_i(\tau,\vec{x})  \hat{B}_i(\tau,\vec{x}).
\eeq
Similarly, for each polarization of canonical tensor fluctuations $Q_\lambda$, we have
\beq\label{SQl}
S\left[\hat{Q}^{(\rm p)}_{\lambda}\right]=\frac{1}{2} \int \d \tau \d^{3} k\left\{\hat{Q}^{(\rm p)\,'}_{\lambda}\hat{Q}^{(\rm p)\,'}_{\lambda}-\left[k^{2}- \fr{a''(\tau)}{a(\tau)}\right] \hat{Q}^{({\rm p})^2}_{\lambda} + 2\hat{Q}^{(\rm p)}_\lambda ~ \hat{J}^{(\rm p)}_\lambda (\tau, \vec{k})\right\},
\eeq
where we have labeled the quantities with $^{(\rm p)}$ to distinguish this primordial contribution from the induced component we study in Appendix \ref{AppC}. In eq. \eqref{SQl}, the primordial source term involving gauge fields is given by the following Fourier transform
\beq\label{JPL}
\hat{J}^{(\rm p)}_{\lambda}(\tau,\vec{k})
 \equiv - \fr{a(\tau)^3}{\Mp}\Pi_{ij,\lambda} (\vec{k})\int \fr{\d^3 x}{(2\pi)^{3/2}} ~{\mathrm e}^{-i\vec{k}.\vec{x}} \bigg[ \hat{E}_i (\tau, \vec{x}) \hat{E}_j (\tau, \vec{x})+ \hat{B}_i (\tau, \vec{x}) \hat{B}_j (\tau, \vec{x})\bigg].
\eeq
Next, we will study the scalar and tensor modes in the presence of vector modes sources, \ie $\hat{J}_\phi$ and $\hat{J}^{(\rm p)}_\lambda$ which we will discuss separately in the following subsections \ref{SS} and \ref{ST}, respectively.
\subsection{Sourced scalar fluctuations}\label{SS}
In the presence of gauge field production, the coupling $\phi F \tilde{F}$ may significantly affect inflaton fluctuations through the inverse decay of amplified fluctuations in the gauge field sector:  $\delta A + \delta A \to \delta \phi$. In order to investigate these effects, we will focus on the mode equation of the canonical variable $\hat{Q}_\phi = a \delta \phi$, which can be derived from \eqref{SQp} as  
\beq\label{Qphi}
\bigg(\partial^2_\tau + k^2 + m^2_{\mathrm{eff}} (\tau)\bigg)\hat{Q}_\phi(\tau,\vec{k}) = \hat{J}_{\phi}(\tau,\vec{k}) \equiv \fr{\alpha_{\rm c}a^3}{f}\int\fr{\d^3 p}{(2\pi)^{3/2}} \hat{E}_{i}(\tau, \vec{k}-\vec{p}) \hat{B}_{i}(\tau,\vec{p}).
\eeq
 In terms of the slow-roll parameters and background quantities, the time dependent mass term is given by 
 \beq\label{effm}
 m^2_{\mathrm{eff}} (\tau) = - (aH)^2 \left[ 2 - \epsilon + \fr{3\eta}{2} + \fr{1}{4}\eta^2 -\fr{1}{2}\epsilon \eta +\fr{\dot{\eta}}{2H}\right] ,
 \eeq
 where we defined 
 \beq\label{srp}
 \epsilon \equiv \fr{\dot{\phi}^2}{2H^2\Mp^2} , ~~~~~~~~ \eta \equiv \fr{\dot{\epsilon}}{\epsilon H}.
 \eeq
 
As the inflaton traverses steep cliffs, all the terms in $m_{\rm eff}$ experience notable oscillations, including transient violations of the slow-roll parameter,\ie $|\eta| \gtrsim \mathcal{O}(1) $. This situation should be contrasted with vanilla slow-roll inflation where to a good approximation one can safely assume $ m^2_{\mathrm{eff}} \to -2 /\tau^2$. In the model we are considering, we therefore need take into account the effects that this departure from slow-roll regime may imprint on the homogeneous and particular solutions of the eq. \eqref{Qphi}. In this work, for the calculation of vacuum power spectrum, we will use numerical tools that are designed to solve coupled background and fluctuations equations during inflation as we explain further in Section \eqref{Sec4p4p1}.

For the purpose of calculating sourced scalar fluctuations, we first seperate $\hat{Q}_\phi$ in \eqref{Qphi} into a vacuum mode, $\hat{Q}^{(v)}_{\phi}$, \ie solution to the homogeneous part of \eqref{Qphi} and the sourced mode $\hat{Q}^{(s)}_\phi$, \ie particular solution of \eqref{Qphi}. The vacuum mode can be decomposed in the standard way as
\begin{align}\label{VMp}
 \hat{Q}_{\phi}^{(v)}(\tau, \vec{k}) &= Q_{\phi}^{(v)}(\tau, k)~ \hat{a}(\vec{k})+ Q_{\phi}^{(v)*}(\tau, k)~ \hat{a}^{\dagger}(-\vec{k}), 
\end{align}
where $\hat{a}$  and $\hat{a}^{\dagger}$ are the creation and annihilation operators for $\hat{Q}_{\phi}^{(v)}$. The solution for the complex vacuum mode function can be therefore obtained by solving the following equation,
\beq\label{hsQphi}
\bigg(\partial^2_\tau + k^2 + m^2_{\mathrm{eff}} (\tau)\bigg){Q}^{(v)}_\phi(\tau,k) = 0.
\eeq
On the other hand, the particular solution of \eqref{Qphi} is given by 
\beq\label{Qphis}
\hat{Q}_{\phi}^{(s)}(\tau, \vec{k}) = \int^{\tau} d\tau'~ G^{\phi}_k(\tau,\tau')~ \hat{J}_\phi(\tau',\vec{k}),
\eeq
where $ G^{\phi}_k(\tau,\tau')$ is the Green's function associated with the homogeneous part of the eq. \eqref{Qphi} and can be constructed using the solutions of eq. \eqref{hsQphi} as
\beq\label{GFphi}
G_{k}^{\phi}\left(\tau, \tau^{\prime}\right)=i \Theta\left(\tau-\tau^{\prime}\right)\left[Q^{(v)}_\phi(\tau, k)Q^{(v)^*}_\phi(\tau', k)-Q^{(v)^*}_\phi(\tau, k) Q^{(v)}_\phi(\tau', k)\right].
\eeq
As we mentioned earlier, it is not possible to obtain a closed form expression for $ G^{\phi}_k(\tau,\tau')$ in terms of known elementary functions when the background deviates from vanilla slow-roll evolution during inflation, \ie during the times when the inflaton rolls through the steep cliffs in its potential \eqref{Vb}. Nevertheless, in Appendix \ref{AppC}, we will introduce a procedure to simplify the form of $G^{\phi}_k$, allowing for the computation of the sourced scalar correlators we are interested using semi-analytic techniques.

Keeping these in mind, we will use comoving curvature perturbation to calculate scalar correlators. In the spatially flat gauge, it is proportional to the sum of vacuum and sourced inflaton perturbation as
\beq\label{CP}
\hat{\mathcal{R}}(\tau,\vec{k}) = \fr{H}{a\dot{\phi}} \hat{Q}_\phi(\tau,\vec{k}) = \fr{H }{a \dot{\phi}} \left(\hat{Q}_{\phi}^{(v)}(\tau, \vec{k}) + \hat{Q}_{\phi}^{(s)}(\tau, \vec{k})  \right).
\eeq 
Using eq. \eqref{CP}, we provide a detailed derivation of the full power spectrum of curvature perturbation in Appendix \ref{AppC} by taking into account the transient deviation of the background from its slow-roll attractor regime, \ie during the rollover of inflaton $\phi$ through smooth plateaus followed by steep cliff(s) shown in Figure \ref{fig:V}.
\subsection{Sourced tensor fluctuations}\label{ST}
Enhanced vector fields may also influence tensor fluctuations substantially. We can investigate such effects focusing on the action \eqref{SQl}, which we vary to obtain the mode equation for $Q_\lambda$ as
\beq\label{ctme}
\left(\partial^2_\tau + k^2 -\fr{a''(\tau)}{a(\tau)}\right)\hat{Q}^{(\rm p)}_\lambda(\tau,\vec{k}) = \hat{J}^{(\rm p)}_{\lambda}(\tau,\vec{k}).
\eeq

Using the Fourier decomposition of $\vec{E}$ and $\vec{B}$ fields, the source term \eqref{JPL} that appear on the right hand side of \eqref{ctme} can be written as a convolution in momentum space
\beq\label{ts}
\hat{J}^{(\rm p)}_{\lambda}(\tau, \vec{k})=-\frac{a^{3}(\tau)}{M_{\mathrm{pl}}} \Pi_{i j, \lambda}(\vec{k}) \int \frac{\mathrm{d}^{3} p}{(2 \pi)^{3 / 2}}\left[\hat{E}_{i}(\tau, \vec{k}-\vec{p}) \hat{E}_{j}(\tau,\vec{p})+\hat{B}_{i}(\tau,\vec{k}-\vec{p}) \hat{B}_{j}(\tau,\vec{p})\right].
\eeq

Similar to the case for scalar fluctuations, we solve for $Q_\lambda$ in \eqref{ctme}  by separating $Q_\lambda$ into a vacuum mode, $Q^{(v)}_\lambda$ and the sourced mode $Q^{(s)}_\lambda$.  The vacuum mode is given by 
\begin{align}\label{VM}
\nn \hat{Q}_{\lambda}^{(v,{\rm p})}(\tau, \vec{k}) &= Q_{\lambda}(\tau, k)~ \hat{a}_{\lambda}(\vec{k})+ Q_{\lambda}^{*}(\tau, k)~ \hat{a}_{\lambda}^{\dagger}(-\vec{k}), 
\\ Q_{\lambda}(\tau, k) &=\frac{\mathrm{e}^{-i k \tau}}{\sqrt{2 k}}\left(1-\frac{i}{k \tau}\right),
\end{align}
where $\hat{a}_{\lambda}^{\dagger}$ creates a graviton with helicity $2\lambda$. On the other hand, the sourced contribution can be written formally as
\beq\label{SQ}
\hat{Q}_{\lambda}^{(s,{\rm p})}(\tau, \vec{k}) = \int^{\tau} d\tau'~ G_k(\tau,\tau')~ \hat{J}^{(\rm p)}_\lambda(\tau',\vec{k}),
\eeq
where the Green's function can be obtained from the homogeneous part of \eqref{ctme} as
\beq\label{tgf}
G_{k}\left(\tau, \tau^{\prime}\right)=\Theta\left(\tau-\tau^{\prime}\right) \frac{\pi}{2} \sqrt{\tau \tau^{\prime}}\left[J_{3 / 2}(-k \tau) Y_{3 / 2}\left(-k \tau^{\prime}\right)-Y_{3 / 2}(-k \tau) J_{3 / 2}\left(-k \tau^{\prime}\right)\right],
\eeq
where $J_\nu$ and $Y_\nu$ are Bessel functions of real argument. 

{\bf Primordial tensor power spectrum:} The origin of sourced tensor fluctuations in this model is identical to models studied in \cite{Namba:2015gja,Ozsoy:2020ccy}. We will therefore omit a detailed derivation of the sourced primordial tensor power spectrum and refer the interested reader to Appendix D of \cite{Namba:2015gja} or Appendix B of \cite{Ozsoy:2020ccy} for a detailed discussion on this matter. Keeping this in mind, we define the tensor power spectrum as 
\beq\label{DTPS}
\frac{k^{3}}{2 \pi^{2}}\left\langle\hat{h}_{\lambda}(\tau, \vec{k}) \hat{h}_{\lambda^{\prime}}(\tau, \vec{k}^{\prime})\right\rangle \equiv \delta_{\lambda \lambda^{\prime}}\, \delta \left(\vec{k}+\vec{k}^{\prime}\right) \,  \mathcal{P}_{\lambda}(\tau,k).
\eeq
Using the relation between the tensor mode operators $\hat{h}_\lambda$ and the canonical mode $\hat{Q}_\lambda$: 
\beq\label{rqtoh}
\hat{h}_\lambda (\tau, k) = \Pi_{ij, \lambda} (\vec{k}) \hat{h}_{ij} (\tau, \vec{k}) = \fr{2}{\Mp a(\tau)} \hat{Q}_\lambda(\tau, \vec{k}),
\eeq
The total primordial tensor power spectrum can be written as the sum of uncorrelated vacuum and sourced part which can be obtained using \eqref{VM} and \eqref{SQ} as,
\begin{align}\label{ptps}
\mathcal{P}^{(\rm p)}_{\lambda} (k)&= \mathcal{P}^{(v,{\rm p})}_{\lambda} (k) + \mathcal{P}^{(s,{\rm p})}_{\lambda} (k)
\\ &=\frac{H^{2}}{\pi^{2} \Mp^{2}} +\frac{H^{4}}{64 \pi^{4} \Mp^{4}} f_{2, \lambda}\left(\xi_{*}, x_{*}, \delta\right),
\end{align}
where the function that parametrizes the gauge field production is given by \cite{Ozsoy:2020ccy}
\begin{align}\label{f2lalt}
\nn  f_{2,\lambda} \left(\xi_*,\fr{k}{k_*},\delta\right) &=\fr{1}{4} \int_{1}^{\infty} \d x \int_{0}^{1} \d y \, \fr{(1-y^2)^2\,(1-\lambda x)^4}{\sqrt{x^2-y^2}}  N^{2}\bigg(\xi_{*}, \fr{x-y}{2} x_{*}, \delta\bigg) N^{2}\bigg(\xi_{*}, \fr{x+y}{2} x_{*}, \delta\bigg)  \\\nn\\
&\quad\quad\quad\quad\quad\quad\quad\quad\quad \times   \mathcal{I}^{2}\left[\xi_{*}, x_{*}, \delta, \fr{x+y}{2}, \fr{x-y}{2}\right],
\end{align}
where $x_* = - k\tau_* = k/k_*$ denoting the ratio of the physical momentum to the horizon side at the time when $\xi$ reaches its peak value $\xi_* = \alpha_{\rm c} \delta$ (See \eg eq. \eqref{xiapp}).
Recall that \eqref{f2lalt} involves the normalization factors $N$ of gauge fields which we derived in Appendix \ref{AppA} and the function $\mathcal{I}$ is defined as \cite{Ozsoy:2020ccy}
\beq
\mathcal{I}\bigg[\xi_{*}, x_{*}, \delta, \tilde{p}, \tilde{q}\bigg] \equiv \mathcal{I}_{1}\bigg[\xi_{*}, x_{*}, \delta, \sqrt{\tilde{p}}+\sqrt{\tilde{q}}\bigg]+\frac{\sqrt{\tilde{p} \tilde{q}}}{2} \mathcal{I}_{2}\bigg[\xi_{*}, x_{*}, \delta, \sqrt{\tilde{p}}+\sqrt{\tilde{q}}\bigg]
\eeq
with $\mathcal{I}_1$ and $\mathcal{I}_2$ representing the time integral of the gauge field sources. They are defined as
\begin{align}
&\mathcal{I}_{1}\bigg[\xi_{*}, x_{*}, \delta, Q\bigg] \equiv \int_{0}^{\infty} \d x^{\prime}\left(x^{\prime} \cos x^{\prime}-\sin x^{\prime}\right) \sqrt{\frac{\xi\left(x^{\prime}\right)}{x^{\prime}}} \exp \left[-\frac{2 \sqrt{2\xi_{*}}}{\delta}\frac{x'^{1/2}}{|\ln(x'/x_*)\,|} Q\right],\\
&\mathcal{I}_{2}\bigg[\xi_{*}, x_{*}, \delta, Q\bigg] \equiv \int_{0}^{\infty} \d x^{\prime}\left(x^{\prime} \cos x^{\prime}-\sin x^{\prime}\right) \sqrt{\frac{x^{\prime}}{\xi\left(x^{\prime}\right)}} \exp \left[-\frac{2 \sqrt{2\xi_{*}}}{\delta}\frac{x'^{1/2}}{|\ln(x'/x_*)\,|} Q\right].
\end{align}

\subsection{Summary of the primordial power spectra}\label{Sec3p3}
In this work we are interested in the 2-pt correlators of scalar and tensor fluctuations which are defined as in eqs. \eqref{DRPS} and \eqref{DTPS}. As we discussed previously, the primordial power spectra can be written as a sum of uncorrelated contributions of quantum vacuum fluctuations and those sourced by the vector fields:
\begin{align}
 \nn\mathcal{P}_{\mathcal{R}}(k)& =  \mathcal{P}^{(v)}_{\mathcal{R}}(k) +  \mathcal{P}^{(s)}_{\mathcal{R}}(k), ~~~   \mathcal{P}^{(\rm p)}_{\lambda}(k) =  \mathcal{P}^{(v,{\rm p})}_{\lambda}(k) +  \mathcal{P}^{(s,{\rm p})}_{\lambda}(k).
 \end{align}
In this model, in contrast to the vacuum fluctuations of the metric, only $-$ helicity states of the sourced tensor fluctuations are amplified in the presence of vector field sources $A_-$. Therefore, in the rest of this work, we will only consider $\mathcal{P}^{(s,{\rm p})}_{-}$ to study the phenomenology of the rolling bumpy axion model. Following our discussion in the previous section, the vacuum power spectra is given by the following expressions
\begin{align}\label{PSv}
 \mathcal{P}^{(v)}_{\mathcal{R}}(k) &= \lim_{\tau \to 0^{-}} \fr{k^3}{2\pi^2} \left(\fr{ H }{a\dot{\phi}}\right)^2 \big|Q^{(v)}_\phi(\tau,\vec{k})\big|^2 , ~~~   \mathcal{P}^{(v,{\rm p})}_{\lambda}(k) = \fr{H^2}{\pi^2 \Mp^2} .
 \end{align}
Focusing on a representative example of background evolution in the model we described above, we will compute the scale dependence of vacuum power spectra in \eqref{PSv} numerically in Section \ref{Sec4}. On the other hand, we will compute the sourced contributions to the scalar and tensor power spectrum using the formulas we developed in Appendix \ref{AppC} and in Section \ref{ST}. These contributions can be summarized as
\beq\label{PSs}
 \mathcal{P}^{(s)}_{\mathcal{R}}(k) =  \mathcal{P}^{(v)}_{\mathcal{R}}(k) \fr{H^2}{64\pi^2 \Mp^2} f_{2,\mathcal{R}}(\xi_*,x_*,\delta),~~~~\mathcal{P}_{-}^{(s,{\rm p})}(k) \simeq \fr{H^4}{64 \pi^4 \Mp^4}\, f_{2,-} (\xi_*,x_*,\delta),
\eeq

where dimensionless functions $f_{i,j}$ with $i = 2$ and $j = \{\mathcal{R},-\}$ parametrize the dependence of the sourced power spectra on the background model. In particular,  during the roll of the axion on a cliff like region of its wiggly potential, the effective coupling $\xi$ between the vector fields and $\phi$ increases, leading to a bump in $\xi$ in time direction. During the time where $\xi$ reaches its peak value $\xi_*$, the amplification of the gauge field modes that crosses the horizon is maximal, leading to the efficient enhancement of certain range of wave-numbers localized in momentum space. For the power spectra of $\mathcal{R}^{(s)}$ and $h^{(s,{\rm p})}_{-}$ sourced by the vector fields, this directly translates into a localized bump in momentum space controlled by the ratio $x_* = k/k_*$. The height of this scale dependent signal depends on the maximum value $\xi_*$ achieved by $\xi$ whereas the width depends on the number of e-folds $\dot{\phi}$ (or $\xi$) significantly differs from zero: $\Delta N \simeq \delta^{-1}$, manifesting the dependence of the signal on the ratio $\delta \propto m/H$. For larger $\delta$, $\dot{\phi}$ reaches its peak faster before it reduces to very small values in the plateau regions of its potential \eqref{Vb}. In this case, the roll of $\phi$ influences fewer modes of the gauge fields, reducing the width of the bump in the power spectra.

For a fixed $\xi_*$ and $\delta$, we found that the momentum dependence (\ie $x_* = k/k_*$) of the dimensionless functions $f_{i,j}$ can be described by a log-normal shape \cite{Namba:2015gja,Ozsoy:2020ccy},
\beq\label{fpheno}
f_{i, j}\left(\frac{k}{k_{*}}, \xi_{*}, \delta\right) \simeq f_{i, j}^{c}\left[\xi_{*}, \delta\right] \exp \left[-\frac{1}{2 \sigma_{i, j}^{2}\left[\xi_{*}, \delta\right]} \ln ^{2}\left(\frac{k}{k_{*}\, x_{i, j}^{c}\left[\xi_{*}, \delta\right]}\right)\right].
\eeq

As suggested by the expression \eqref{fpheno}, the information about the location, width and the height of the sourced signals in \eqref{PSs} depends on the motion of $\phi$ in the step-like features of its wiggly potential, particularly through $\xi_*$ and $\delta$ dependence of the functions $x^c_{i,j}, \sgm_{i,j}, f^c_{i,j}$. It is clear from \eqref{fpheno} that $f_{i,j}$ is maximal at $k = k_* \, x^c_{i,j}$, where it evaluates to $f^c_{i,j}$ whereas $\sgm_{i,j}$ controls the width of the signal. In the next section, focusing on a representative background model of bumpy axion inflation (with a fixed $\delta$), we will derive accurate formulas for these functions by fitting the right hand side of eq. \eqref{fpheno} to reproduce the position, height and width of the sourced signal parametrized within the integrals of $f_{i,j}$ defined in Section \ref{ST} and Appendix \ref{AppC}.
\section{Phenomenology of the bumpy axion inflation}\label{Sec4}
The motion of an initially displaced $\phi$ around the plateau like regions is expected to be smooth and slowly varying due to the small slopes the scalar potential exhibits (See \eg Figure \ref{fig:V}). Therefore, plateau like regions are suitable to sustain the inflationary dynamics required to produce nearly scale invariant scalar fluctuations at CMB scales. On the other hand, the roll of the inflaton $\phi$ through the cliff(s) of its wiggly potential leads to efficient production of gauge field fluctuations that can be considered as a source of curvature and metric perturbations through the corresponding inverse decay processes: $\delta A_{-}\, + \,\delta A_{-} \to \delta \phi$ and $\delta A_{-} \,+\, \delta A_{-} \to h_-$. In the presence of the coupling \eqref{LINT}, we studied the influence of such additional channels on scalar and tensor fluctuations in Section \ref{SS} and \ref{ST}. The main structure of the resulting 2-pt correlators sourced by gauge fields is discussed in Section \ref{Sec3p3} and are given by eq. \eqref{PSs}. In this section and the following subsections, we will focus on a typical background model of axion inflation in the bumpy regime ($\beta \lesssim  1$) to study the phenomenological implications of this model at CMB and sub-CMB scales.

For this purpose, we consider a representative background model of axion inflation by focusing on the following parameter choices in the bumpy regime of scalar potential in eq. \eqref{Vb},
\beq\label{PC}
\beta \equiv \fr{\Lambda^{4}}{m^2 f^2}=0.9958,~~~~~\alpha\equiv \fr{\Mp}{f}= 3.3.
\eeq
We note that for $\Mp/f \sim \mathcal{O}(1)$ and an intermediate field range $\Delta \phi \sim 3 \Mp$ we consider in this work, $\beta$ should be tuned as in eq. \eqref{PC} to obtain sufficient amount of e-folds during inflation. In general, decreasing $f$ \footnote{In the context of axion inflation, see \cite{Kobayashi:2010pz,Flauger:2014ana,Kobayashi:2015aaa,Kappl:2015esy,Choi:2015aem} for models that adopts parametrically smaller $f$. For example, in the axion monodromy model $f \sim 10^{-2} - 10^{-6} \Mp$ which leads to resonances in perturbations resulting with oscillations in the spectral index $n_s$ \cite{Flauger:2014ana}. On the other hand, axions in string theory tend to have decay constants between the GUT and Planck scale \cite{Svrcek:2006yi}. } with respect to $\Mp$ (increasing $\alpha$) increases the number of wiggles in the scalar potential at a given field range and so the required tuning of $\beta$ in the bumpy regime $\beta \lesssim 1$ or the necessary field range without tuning $\beta$ to obtain enough inflation \cite{Parameswaran:2016qqq}. 
\subsection{Bumpy rides during inflation: slow roll - fast roll - slow roll}\label{Sec4p2}
Assuming negligible backreaction (See Appendix \ref{AppE}) from gauge fields, we now study the inflationary evolution on a flat FRW background. We focus on the parameters choices given by eq. \eqref{PC} in  the axion potential \eqref{Vb} to study the following background equations:
\begin{align}\label{BE}
\nn H^{2}=& \frac{V(\phi)}{M_{\mathrm{pl}}^{2}(3-\epsilon)} \\
\frac{\mathrm{d}^{2} \phi}{\mathrm{d} N^{2}}&+(3-\epsilon) \frac{\mathrm{d} \phi}{\mathrm{d} N}+\frac{1}{H^{2}} V^{\prime}(\phi)=0
\end{align}

where $\d N = \d \ln a(t)$ and the Hubble slow-roll parameter is defined by $\epsilon\equiv-\dot{H}/{H^{2}}=(2 M_{\mathrm{pl}}^{2})^{-1}\left(\mathrm{d} \phi/{\mathrm{d} N}\right)^{2}$. We numerically solve the set of equations in eq. \eqref{BE} assuming initially the system is in the slow-roll attractor regime, defined by the condition ${\mathrm{d} \phi}/{\mathrm{d} N}=-{V^{\prime}(\phi)}/{V(\phi)}.$
\begin{figure}[t!]
\begin{center}
\includegraphics[scale=0.6]{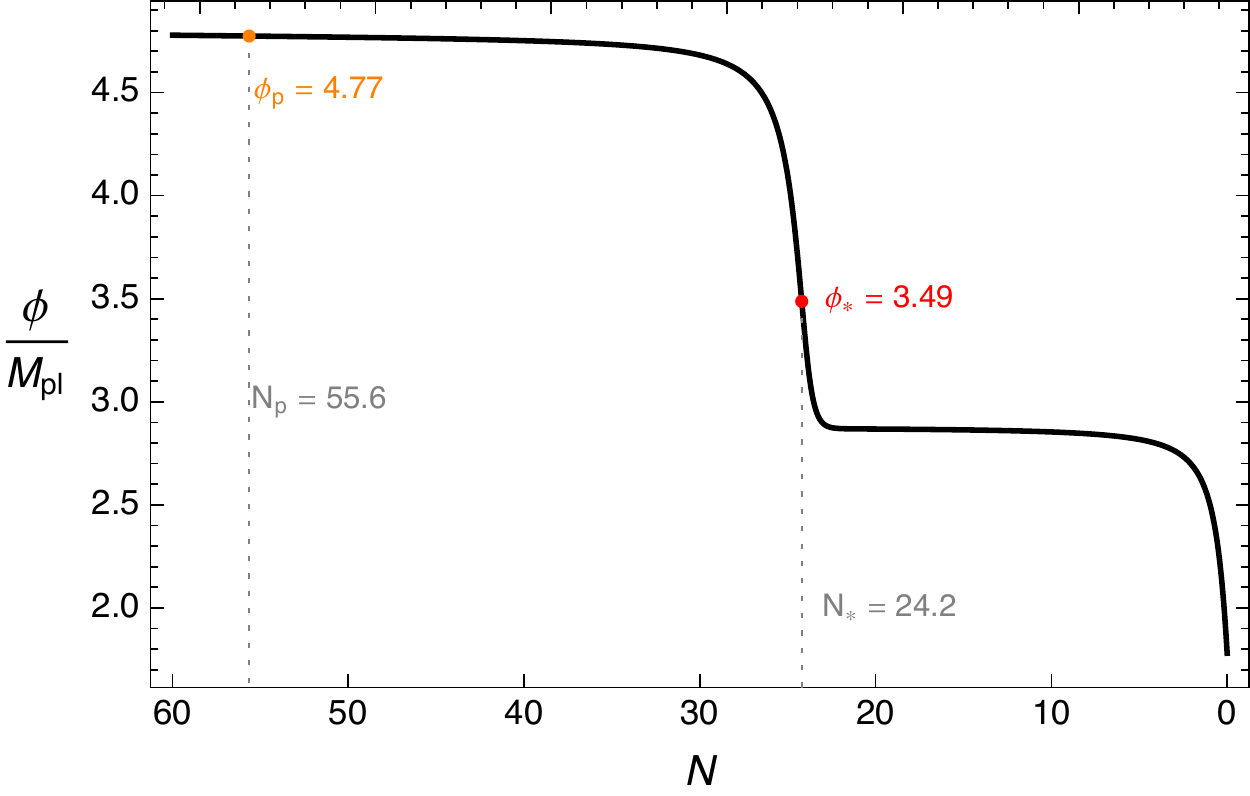}\includegraphics[scale=0.67]{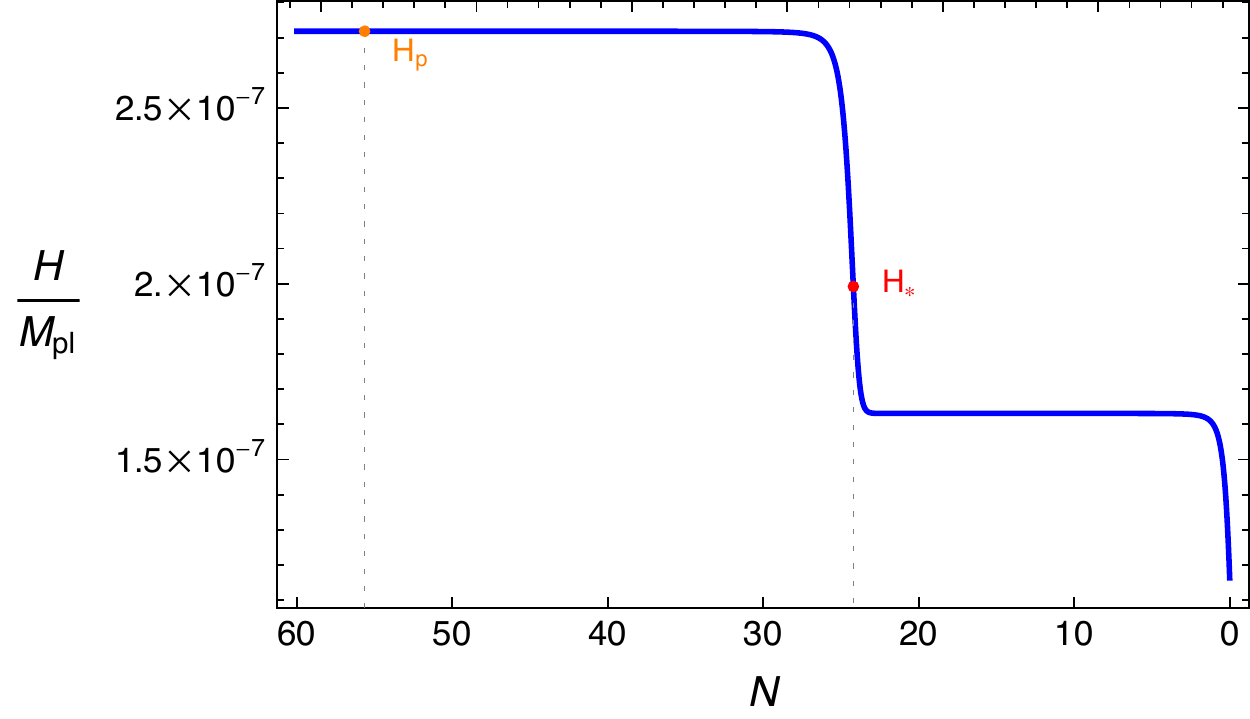}
\end{center}
\caption{The evolution of $\phi$ (left) and Hubble parameter $H$ (right) with respect to e-folds for the parameter choices given by \eqref{PC} (See also Table \ref{tab:bparams}) in the potential \eqref{Vb}. \label{fig:phiandH}}
\end{figure}
\begin{figure}[t!]
\begin{center}
\includegraphics[scale=0.64]{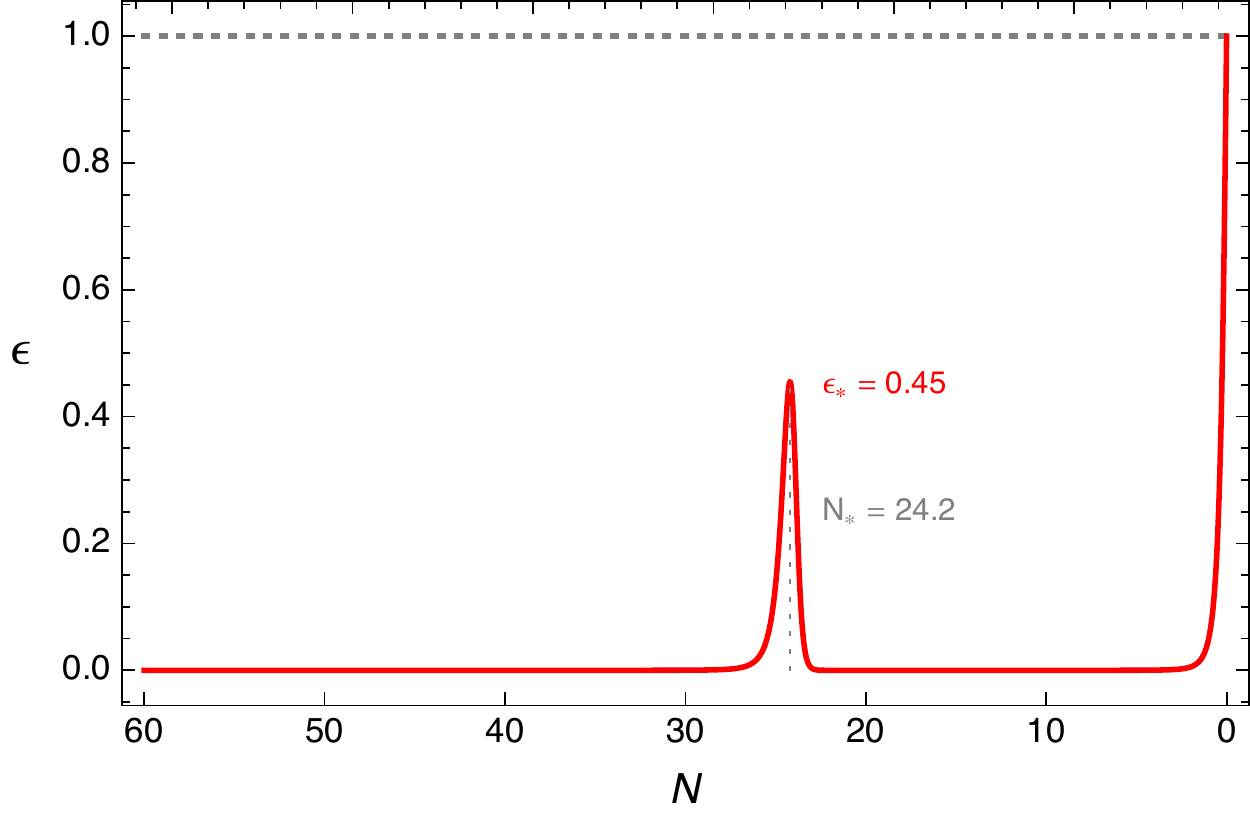}\includegraphics[scale=0.64]{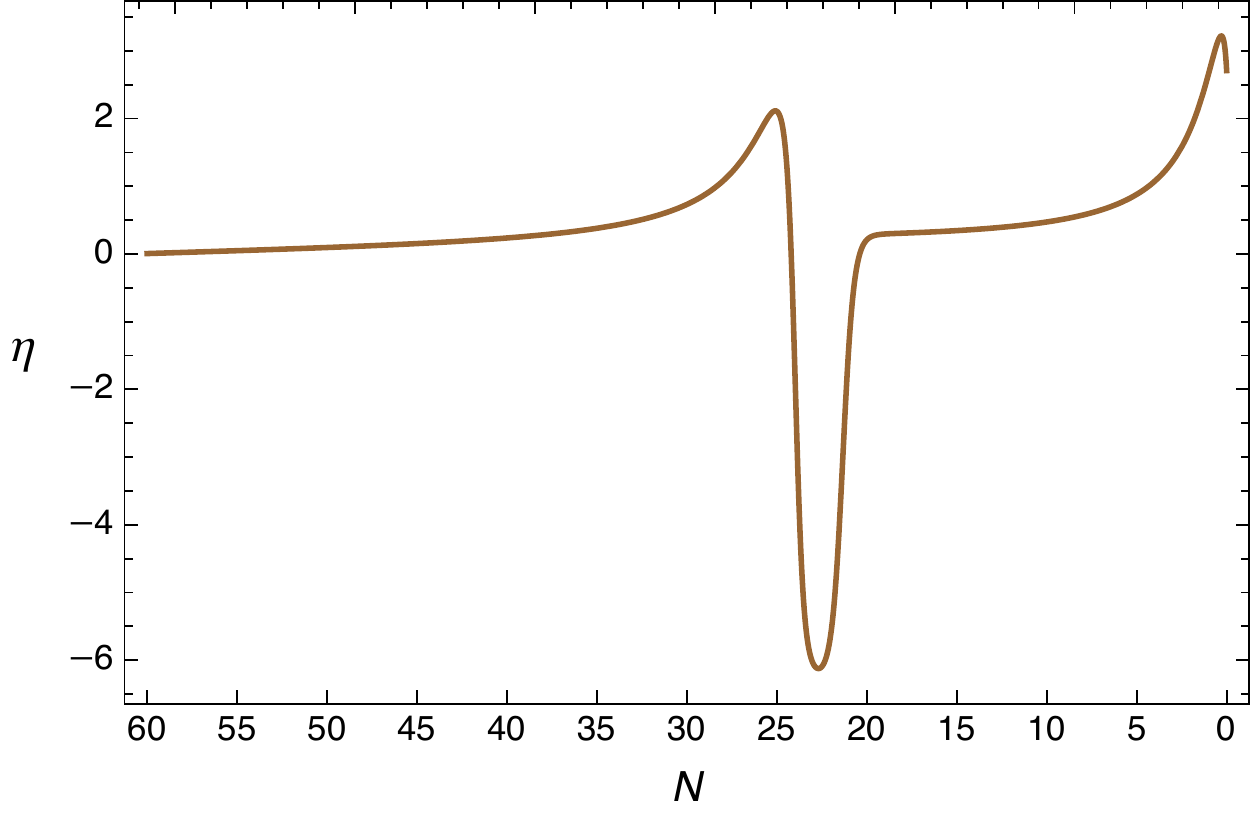}
\end{center}
\caption{The evolution of slow-roll parameters $\epsilon$ (left) and $\eta$ (right) with respect to e-folds during inflation for the same parameter choice provided in Figure \ref{fig:phiandH}. \label{fig:epsandeta}}
\end{figure} 
\indent In Figure \ref{fig:phiandH}, we present the resulting field profile $\phi$ and Hubble rate $H$ as a function of e-folds during inflation where we set $\phi =  4.8 \,\Mp$ initially. We observe that the inflaton slowly rolls down the smooth plateau-like regions, sustaining an almost constant Hubble friction. However, whenever it meets a cliff, $\phi$ speeds up quickly, until it reaches the next plateau where Hubble friction rapidly slows it right back down again. The system is in a slow-roll attractor regime within the plateau, but departs from it during the acceleration/fast roll through the steeper cliff and the during the deceleration when rolling into the next flat plateau following the steep cliff. This behaviour can be seen clearly from Figure \ref{fig:epsandeta} where we the evolution of the slow-roll parameter $\epsilon$ and $\eta \equiv \dot{\epsilon}/\epsilon H$ with respect to e-folds $N$ is shown. We see that $\epsilon$ peaks as $\phi$ accelerates ($\eta > 0$) down the steep cliffs  and then $\epsilon$ reduces back again as $\phi$ decelerates ($\eta < 0$) into the plateaus. 

During the time where $\epsilon$ peaks, the effective coupling between the gauge fields and the axion will be maximal $\xi \propto \sqrt{\epsilon}$, leading to efficient particle production in the gauge field sector (See Section \ref{Sec2p2}). In Section \ref{Sec4p4}, we will study the resulting sub-CMB phenomenology for the scalar and tensor fluctuations at scales during such particle production processes. However, before we proceed, we need to make sure that the predictions of our model are in agreement with the observations at the CMB scales. This will be the topic of the following subsection.

It should be noted that in order to fix the overall scale of the potential and thus the Hubble rate with respect to Planck scale $\Mp$ (See Figure \ref{fig:phiandH}), we need to determine the mass scale $m$ in the scalar potential \eqref{Vb}. For this purpose, first we found that the pivot scale $k_p = 0.05\, {\rm Mpc^{-1}}$ exits the horizon at $N_p \simeq 55.6$ e-folds in the model under considration. We then utilize the normalization of the scalar power spectrum at the pivot scale $\mathcal{P}_\mathcal{R}(k_p) \simeq 2.1\times 10^{-9}$ to fix the overall mass scale $m$ which in turn allows us to determine $\Lambda$ for the given $\beta$ in eq. \eqref{PC}. In this way, we summarize the model parameters that give rise to the background evolution we presented in Figure \ref{fig:phiandH} and \ref{fig:epsandeta} in Table \ref{tab:bparams}.
\subsection{CMB Phenomenology}\label{Sec4p3}
In the previous subsection, we have seen that the slow-roll parameters undergo large oscillations when the field rolls down the steep cliffs and into the plateaus of the potential \eqref{Vb}. However, during the short range of e-folds that is associated with CMB scales, \ie the roll of the scalar field in the first plateau region,  the slow-roll parameters are small and are evolving smoothly. 
\begin{table}[t!]
\begin{center}
\begin{tabular}{ l c }
\hline
\hline
 \cellcolor[gray]{0.9}  ~~&~~    \cellcolor[gray]{0.9} 
   $\Mp/f =3.3$  \\
\hline
$N_p $ ~~&~~ $~~~~~~~55.6$  \\
$m $ ~~&~~ $~~~~~~~1.399\times 10^{-7}\, \Mp$  \\
$\Lambda$ ~~&~~~~~~~~~ $2.057\times 10^{-4}\, \Mp$  \\

\hline
\hline
\end{tabular}
\caption{\label{tab:bparams} The number of e-folds $N_p$ at which the pivot scale crosses the horizon during inflation and the relevant mass scales in the axion potential \eqref{Vb}.}
\end{center}			
\end{table}
\begin{table}[t!]
\begin{center}
\begin{tabular}{ l c c }
\hline
\hline
 \cellcolor[gray]{0.9}
 Observables ~~&~~~  \cellcolor[gray]{0.9}   Case 1: $\Mp/f =3.3$ \\

\hline
$~~~~~~n_s$ ~~&~~~~ $0.9640$ \\
$~~~~~~\alpha_s$ ~~&~~ $-0.0085$ \\
$~~~~~~\beta_s$ ~~&~~ $~~~~-1.1 \times 10^{-4}$ \\
$~~~~~~r$ ~~&~~  $~~~~7.1 \times 10^{-6}$ \\
$~~~~~~n_t$ ~~&~~  $~~-1.2 \times 10^{-6}$ \\

\hline
\hline
\end{tabular}
\caption{\label{tab:bobs} CMB observables in bumpy axion inflation evaluated at the pivot scale $k_p =0.05~ {\rm Mpc^{-1}}$.}
\end{center}			
\end{table}

In order to accurately capture the predictions of the model at CMB scales, we use the model parameters in Table \ref{tab:bparams} and utilize $\mathsf{MultiModeCode}$\footnote{Web page: $\mathsf{www.modecode.org}$.} which is suitable for numerically studying background and perturbation equations when there are large deviations from slow-roll conditions \cite{Mortonson:2010er,Easther:2011yq,Norena:2012rs,Easther:2013rva,Price:2014ufa,Price:2014xpa}. In this way, we determine inflationary observables such as spectral index of scalar fluctuations $n_s$, its running $\alpha_s$, its running of the running $\beta_s$, tensor-to-scalar ratio $r$ and spectral index of tensor fluctuations at the pivot scale. We list these observables in Table \ref{tab:bobs} which shows agreement with the recent Planck data\footnote{In the model we consider since the running of the running $\beta_s$ is two orders of magnitude below $\alpha_s$ we will not consider Planck results including $\beta_s$.} (TT,TE,EE+lowE+lensing+BK15) \cite{Akrami:2018odb} at $k_p = 0.05\, {\rm Mpc^{-1}}$:
\bea\label{P18}
\nn n_s &=& 0.9639\pm 0.0044,\\
 \alpha_s &=& -0.0069\pm0.0069, ~~~~~r < 0.067.
\eea
From Table \ref{tab:bobs}, we observe that the model exhibit a mild running $\alpha_s$ at CMB scales, which is a typical feature of wiggly potentials \cite{Kobayashi:2010pz, Parameswaran:2016qqq}. On the other hand, the existence of flat plateau like regions in the potential leads to the required amount of inflation generically for intermediate field excursion in Planck units: in the example we present in this section, we have $\Delta \phi = 2.99\, \Mp$ between the time pivot scale exits the horizon and the end of inflation where $\epsilon = 1$. As the CMB scales exit the horizon while the axion rolls on the flat plateau region of the potential where  $V'(\phi) \to 0$, the model also exhibits a small tensor-to-scalar ratio $r \approx 10^{-5}$ while the tensor power spectrum obtains a tiny red tilt, $n_t \approx 10^{-6}$. We note that such small values of $r$ is beyond the reach of future CMB polarization missions such as CMB-S4 \cite{Abazajian:2016yjj} and LiteBIRD \cite{Hazumi:2019lys}. It is also worth mentioning that due to the pronounced axion modulations ($\propto \phi$) in the scalar potential, sizeable running of the spectral index $\alpha_s$ typically restricts the choice of $N_p$ allowed by CMB observations (See \eg eq. \eqref{P18}) to be within $1 \%$ of the value we provide in Table \ref{tab:bparams}. In the context of axion monodromy, a simple way out of this problem can be obtained by considering a drift factor that is exponentially sensitive to the axion field value $\phi$ \cite{Ballesteros:2019hus} such that modulations are negligible compared to the monomial term in the scalar potential for field values where CMB scales exit the horizon. 

Although the model we consider leads to unobservable tensor fluctuations at CMB scales, in the presence of the coupling in eq. \eqref{LINT}, the fast roll of the axion offers a rich phenomenology in terms of tensor and scalar fluctuations for modes that exits the horizon around $N_* \simeq 24$ (See \eg Figure \ref{fig:phiandH} and \ref{fig:epsandeta}). In the following subsection, we will therefore focus on the sub-CMB phenomenology of the model we introduced in this section.

\subsection{Phenomenology at sub-CMB scales}\label{Sec4p4}
As we discussed in the beginning of Section \ref{Sec4}, the roll of the axion in the step like feature of its potential leads to an additional primordial component of SGWB ($\delta A_{-} + \delta A_{-} \to h_{-}$) that exhibit a peak around $k \sim k_* = a_*H_*$, corresponding to the scales that exit horizon at around $N_* = 24$ in the specific model we studied above. 
The gauge field amplification that produces this primordial SBGW also enhances the scalar perturbations at the corresponding scales ($\delta A_{-} + \delta A_{-} \to \mathcal{R}$). These amplified scalar fluctuations can later lead to a population of PBHs when the corresponding scales re-enter the horizon during radiation dominated universe (RDU). On the other hand, the enhancement of the scalar perturbations required to produce PBH during RDU can also induce significant amount of GWs\footnote{In the post-inflationary universe, the production of SGWB in this way can also occur in alternative cosmological backgrounds, see \eg \cite{Domenech:2019quo} for a study of induced GWs in a cosmological fluid that exhibit a general constant equation of state. In this context, induced GWs can be considered as a probe of thermal history of the universe \cite{Domenech:2020kqm}.} through the coupling of scalar and tensor modes at second-order in perturbation theory \cite{Kohri:2018awv}. In the presence of gauge field sources, this channel can be schematically described as $\delta A_{-} + \delta A_{-}+\delta A_{-} + \delta A_{-} \to \mathcal{R} + \mathcal{R} \to h_{\pm} $ by noting that scalar perturbations do not discriminate between different helicity states of metric perturbation. In the bumpy axion model we are considering here, we review this important component that contributes to the SGWB in Appendix \ref{AppCC}. In this subsection, our main objective is to study the prospects of generating a large population of PBHs together with primordial + induced SBGW that may originate in the bumpy axion inflation for the background evolution we presented in Section \ref{Sec4p2}.

\begin{table}
\begin{center}
\begin{tabular}{| c | c | c | c |}
\hline
\hline
\cellcolor[gray]{0.9}$\{i,j\}$&\cellcolor[gray]{0.9}$\ln(|f^c_{i,j}|) \simeq $&\cellcolor[gray]{0.9}$x^c_{i,j} \simeq $ &\cellcolor[gray]{0.9}$\sgm_{i,j} \simeq $ \\
\hline
\cellcolor[gray]{0.9}$\{2,\mathcal{R}\}$&$ -6.94 + 5.50\,\xi_*- 0.010\,\xi_*^2$ & $2.19 +0.592\, \xi_* + 0.0041\, \xi_* ^2$&$0.447 -0.0065\, \xi_* + 0.00024\, \xi_* ^2$ \\\hline
\cellcolor[gray]{0.9}$\{2,-\}$ & $-7.79 +5.17\, \xi_* - 0.002\, \xi_* ^2$ & $2.51 + 0.963\,\xi_* + 0.0054\, \xi_*^2$&$0.406 -0.0213\,\xi_* + 0.00061\,\xi_* ^2 $\\\hline
\hline
\end{tabular}
\caption{\label{tab:fit1} $\xi_*$ dependence of the height $f^{c}_{i,j}$, location $x^c_{i,j}$ and width $\sgm_{i,j}$ of eq. \eqref{fpheno} for $\delta = 1.57$.}
\end{center}			
\end{table}

For this purpose, we need to calculate sourced contributions to the primordial power spectra in eq. \eqref{PSs} and therefore require the functions $f_{2,\mathcal{R}}$ and $f_{2,-}$ given by eq. \eqref{fpheno} we introduced earlier. To determine the height, width and the location of the peaks in these functions, we use $\delta = 1.57$\footnote{The peak value of the slow-roll parameter $\epsilon_* = 0.453$ shown in Figure \ref{fig:epsandeta} fixes the choice of $\delta = 1.57$ for $\alpha = \Mp/f = 3.3$. This is because $\epsilon_* = 2\delta^2 / \alpha^2$ as can be inferred from \eqref{srp} using \eqref{dphi}.} as implied by the model we study in Section \ref{Sec4p2}. For this parameter choice, we studied the integrals defined in eqs. \eqref{f2lalt} and \eqref{f2RF} for different $\xi_*$ values. In this way, we find that the functions  $x^c_{i,j}, \sgm_{i,j}, f^c_{i,j}$  can be described by smooth second order polynomials within the interval $8.5 \leq \xi_* \leq 12$ which we present in Table \ref{tab:fit1}.

\subsubsection{Assisted PBH production in bumpy axion inflation}\label{Sec4p4p1}
On scales much smaller compared to the CMB probes, the limits on the PBH abundance put an upper bound on the primordial scalar perturbations as the formation of such objects require enhanced scalar fluctuations. In describing the constraints on scalar power spectrum on various sub-CMB scales from PBH abundance, we will mainly follow the limits considered in \cite{Garc_a_Bellido_2017} including effects induced by black hole evaporation \cite{Khlopov:2008qy,Carr:2009jm,Carr:2020gox}, capture of primordial black holes by stars  during its formation \cite{Capela:2014ita}, micro-lensing \cite{Alcock:1998fx,Tisserand:2006zx}, wide binary disruption \cite{Quinn:2009zg} and finally dragging of halo objects into the Galactic nucleus by dynamical friction \cite{Carr:1997cn} (See also \cite{Carr:2016drx}). We would like to point out that there are large astrophysical uncertainties regarding the star formation constraints \cite{Kawasaki:2016pql}, and for this reason we will not include them in our analysis below, where we assume that the corresponding mass window, $10^{20} \lesssim M_{\rm PBH} \,[{\rm g}]\lesssim 10^{22}$, can be compatible with PBH being a significant fraction, or the totality of the dark matter abundance. 

In the bumpy axion inflation model we are focusing, the total primordial power spectrum of curvature perturbation is given by 

\beq\label{Fps}
\mathcal{P}_{\mathcal{R}}(k) = \mathcal{P}^{(v)}_{\mathcal{R}}(k) \bigg[1 + \fr{H^2}{64\pi^2 \Mp^2} f_{2,\mathcal{R}}\left(\xi_*,\fr{k}{k_*},\delta = 1.57\right)\bigg],
\eeq

\begin{figure}[t!]
\begin{center}
\includegraphics[scale=0.64]{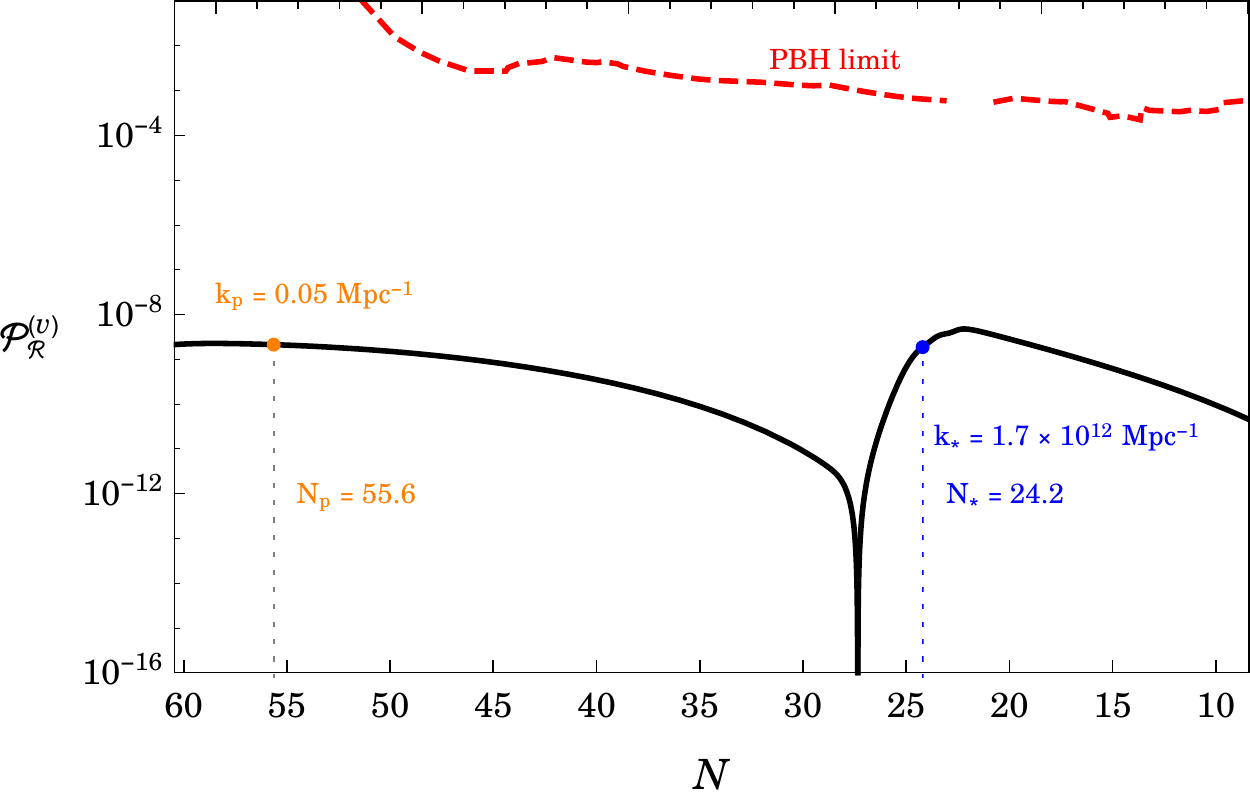}\,\includegraphics[scale=0.64]{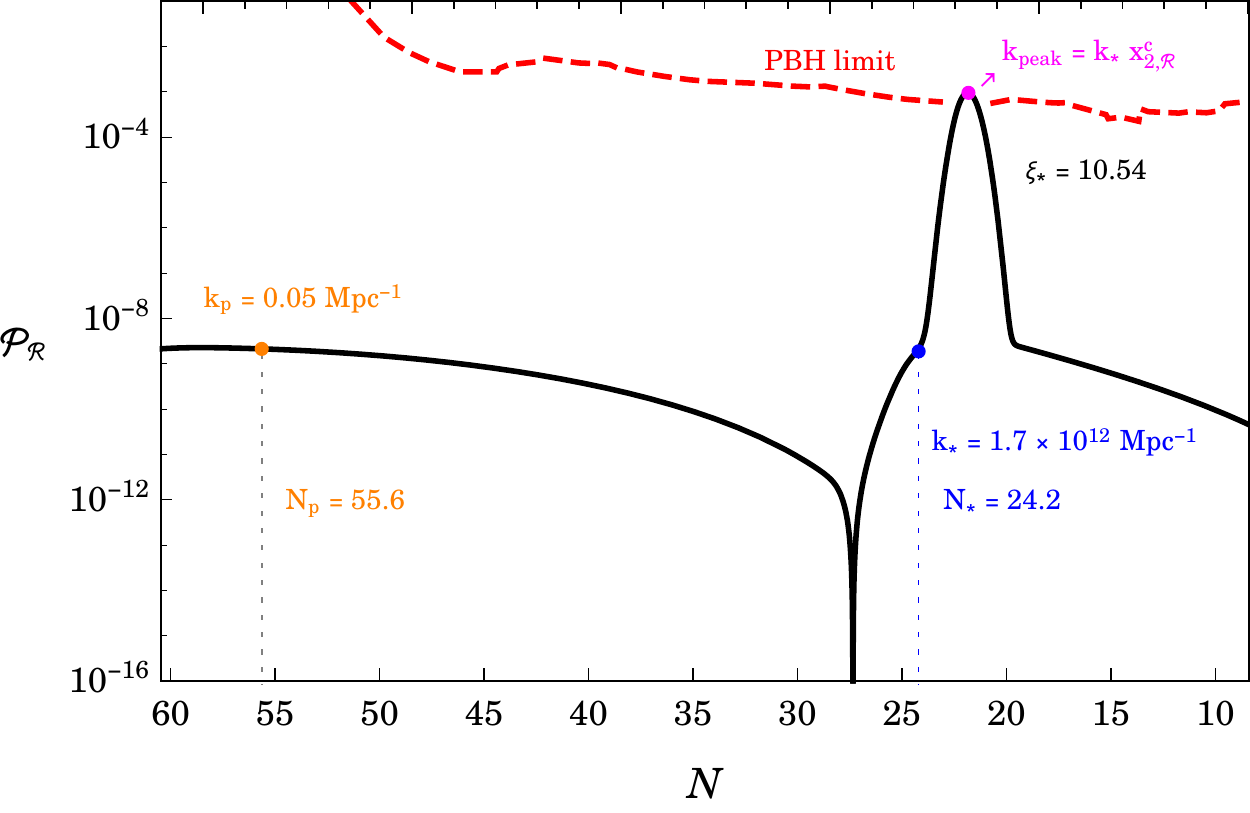}
\end{center}
\caption{The vacuum power spectrum $\mathcal{P}^{(v)}_{\mathcal{R}}$ (left) and the total power spectrum in eq. \eqref{Fps} (right) as a function of number of e-folds in the bumpy axion monodromy model we studied in Section \ref{Sec4p2}. On the right panel, the parameter choice $\xi_* = 10.54$ corresponds to \ie $F_{\rm PBH} = 1$ where PBHs constitutes the total DM abundance.\label{fig:ps}}
\end{figure} 

where the vacuum power spectrum is defined in \eqref{PSv} and $f_{2,\mathcal{R}}$ is given by the shape defined in \eqref{fpheno}. In this expression, given the complexity of background dynamics we studied in Section \ref{Sec4p2}, we calculate the vacuum power spectrum  numerically using $\mathsf{MultiModeCode}$ for the parameter choices provided in Table \ref{tab:bparams}. On the other hand, in order to determine the sourced piece in eq. \eqref{Fps}, we will make use of the background solutions we presented in Section \ref{Sec4p2} together with eq. \eqref{fpheno} and Table \ref{tab:fit1}. In this way, we present the scale dependence of the vacuum and full power spectrum in Figure \ref{fig:ps}, where we replaced the $k$ dependence to number of e-folds using horizon crossing condition for each mode: $k_N = a(N) H(N)$. From the left panel, we realize the characteristic dip in the vacuum scalar power spectrum that is observed for modes that exit the horizon before the system enters the short non-slow roll phase with $\eta < 0$ (\ie before $\epsilon$ reaches its peak value $\epsilon_*$, see \eg Figure \ref{fig:epsandeta}) \cite{Byrnes:2018txb,Carrilho:2019oqg,Ozsoy:2019lyy,Liu:2020oqe}. Following the scales corresponding to the dip, the power in the curvature spectrum first rises due to the short non-attractor phase where $\eta \lesssim -6$ and then decays with a red tilt collectively for modes that exit the horizon during non-attractor and final slow-roll attractor phase. The duration of the non-attractor phase and hence power attained at the peak following the dip depends very sensitively on the parameters of the axion potential $\{m,\Lambda\}$. For the parameter choices\footnote{We have checked that in the vicinity of $\Mp/f = 3.3$, further fine tuning of $\{m,\Lambda\}$ does not lead to enough enhancement (at the order of $10^{7}$) in the vacuum power spectrum required for PBH formation.} we have made in Table \ref{tab:bparams}, we found an order of magnitude growth with respect to CMB scales in the vacuum power spectrum (See Figure \ref{fig:ps}). On the other hand, the presence of gauge field sources leads to an exponential amplification parametrized by $f_{2,\mathcal{R}}$ factor in the total curvature power spectrum in eq. \eqref{Fps}. In other words, around the time when axion velocity reaches its peak, the exponential amplification of the gauge fields efficiently sources the curvature perturbation via $\delta A_{-} + \delta A_{-} \to \mathcal{R}$. As a result, total curvature power spectrum peaks at scales corresponding to $k_{\rm peak} = k_*\, x^c_{2,\mathcal{R}}$ as presented in the right panel of  Figure \ref{fig:ps}. 

{\bf \underline{PBHs as dark matter:}} In the post-inflationary universe, modes corresponding to the peak of the scalar power spectrum ($k \simeq k_{\rm peak} \simeq 1.5 \times 10^{13}\, {\rm Mpc^{-1}}$) can collapse to form PBHs (with $M_{\rm PBH} \simeq 2.2 \times 10^{-13}\,M_{\odot}$) for fluctuations that posses sufficiently large amplitude. The efficiency of PBH formation depends strongly on the statistical properties of the primordial curvature perturbation. In the model under consideration, sourced scalar fluctuations originate from the convolution of two Gaussian gauge field modes and hence obey $\chi^2$ statistics
\,\cite{Linde:2012bt}. In this case, the fraction $\beta$ of causal regions collapsing onto primordial black holes is related to power spectrum of curvature perturbation by \cite{Lyth:2012yp,Byrnes:2012yx}
\beq\label{bpbh}
\beta(N) = \operatorname{Erfc}\left(\sqrt{\frac{1}{2}+\frac{\mathcal{R}_{c}}{\sqrt{2 P_{\mathcal{R}}(N)}}}\right),
\eeq
where $\mathcal{R}_c$ is the threshold for collapse\footnote{Recent theoretical and numerical studies indicate that  $\mathcal{R}_c = \mathcal{O}(1)$ \cite{Musco:2004ak,Musco:2008hv,Musco:2012au,Nakama:2013ica,Harada:2013epa}. Moreover, it has been argued that the threshold for collapse is non-universal and depends on the shape of primordial power spectrum \cite{Musco:2018rwt}. See however \cite{Escriva:2019phb} for a formulation that may allow for a universal threshold. Note that the value of $\beta$ is highly sensitive to the choice of $\mathcal{R}_c$ which can be compansated by a change in $\mathcal{P}_{\mathcal{R}}$ to produce the same PBH abundance. In this work, we take $\mathcal{R}_c = 1.3$ by adopting the universal value of density threshold $\delta_c \simeq 0.4$ quoted in \cite{Escriva:2019phb} and using the relation $\mathcal{R}_c = 9/(2\sqrt{2})\delta_c$ between curvature and density threshold \cite{Drees:2011hb,Young:2014ana,Motohashi:2017kbs}. } during radiation dominated universe and ${\rm Erfc}(x) = 1- {\rm Erf}(x)$ is the complementary error function. At the time of their formation (\ie upon horizon entry of modes with $k \sim k_{\rm peak}$), a fraction $\left.\gamma \beta(M(k)) \rho\right|_{k=a_{f} H_{f}}$ of the total energy in the Universe turns into PBHs\footnote{The value of the constant of proportionality $\gamma = 0.2$ is suggested by the analytical model in \cite{Carr:1975qj} for PBHs formed during the radiation dominated era.}. After their formation, $\beta$ grows inversely proportional to the cosmic tempertaure $(\propto a)$ until matter-radiation equality, since PBHs essentially behave as pressureless dust $\left(\rho_{\mathrm{PBH}} \propto a^{-3}\right) .$ Therefore, neglecting secondary effects such as accretion and merger of PBHs, the fraction of PBH abundance in dark matter density today can be determined by a simple red-shifting relation as \cite{Inomata:2017okj,Sasaki:2018dmp},
\begin{align}\label{fpbh}
f_{\rm PBH}(M(N)) \simeq \left(\frac{\beta(M(N))}{2.8 \times 10^{-15}}\right)\left(\frac{\gamma}{0.2}\right)^{3 / 2}\left(\frac{g_{*}\left(T_{f}\right)}{106.75}\right)^{-1 / 4}\left(\frac{M(N)}{2.2 \times 10^{-13} \, M_{\odot}}\right)^{-1 / 2},
\end{align}
where $T_f$ is the temperature of the plasma in the radiation dominated universe at the time of PBH formation and the relation between the mass of the black holes and the number of e-folds during inflation is given by \cite{ Garc_a_Bellido_2016},
\beq\label{mpbh}
\fr{M(N)}{{2.2 \times 10^{-13} M_{\odot}}}\simeq 3\, \gamma\, \frac{10^{-7}\, \mathrm{GeV} \times H_{\mathrm{end}, \mathrm{inf}}}{H(N)^{2}}\, \mathrm{e}^{2 N},
\eeq
where $H_{\rm end,inf}$ denotes the Hubble rate at the end of inflation. The total fractional PBH abundance is then simply given by 
\beq\label{Fpbh}
F_{\rm PBH} = \int \d \ln M\,  f_{\rm PBH} (M) = 2 \int \d N \, \left(1+\epsilon(N)\right)\, f_{\rm PBH} (N),
\eeq 
where the integral should be taken over e-folds during axion inflation for which integrand is peaked, \ie from $N_{\rm max} = 25$ to $N_{\rm min}=19$ where the scalar power spectrum peaks as in Figure \ref{fig:ps}. Using eqs. \eqref{mpbh} and \eqref{bpbh} in eq. \eqref{fpbh}, we found the limiting value of $\xi_* \simeq 10.54$ in the bumpy axion inflation which corresponds to a PBH abundance that can account for the totality of DM density in the universe, \ie $F_{\rm PBH} = 1$ in eq. \eqref{Fpbh}. The corresponding peak in the curvature power spectrum is shown in the right panel of Figure \ref{fig:ps}.

\subsubsection{Primordial and Induced GW background from bumpy axion inflation}\label{Sec4p4p2}
In the inflationary scenario we introduced above, there are two\footnote{Here we ignore the GW background that can be produced by the merging of PBH binaries, since their formation until today \cite{Clesse:2016ajp,Mandic:2016lcn}.} distinct populations of SGWB: 
\begin{enumerate}
\item The GW background that originates from the amplified gauge fields during inflation through the channel: $\delta A_{-} + \delta A_{-} \to h_{-}$ which we study in Section \ref{ST}. We label this contribution as ``primordial''.
\item{The induced GW background that originates from the scalar fluctuations that are enhanced by the gauge fields during inflation. The induced GW signal in this case is associated with the enhanced scalar modes that re-enter the horizon to form PBHs during RDU. We label this contribution as ``induced'' and study its production channel:  $\delta A_{-} + \delta A_{-} + \delta A_{-} + \delta A_{-} \to \mathcal{R} + \mathcal{R} \to h_{\pm}$ in Appendix \ref{AppC}.}
\end{enumerate}
We express the amplitude of the stochastic GW background in terms of the present fractional energy density of GWs per logarithmic wavenumber, \ie $\Omega_{\rm gw}$ (See Appendix \ref{AppD}). In terms of the tensor power spectrum of individual contributions we discussed above, it is given by
\begin{align}\label{toto}
\nn \Omega^{(\rm tot)}_{\rm gw}(\tau_0,k)\, h^2 &= \left( \Omega^{(\rm p)}_{\rm gw}(\tau_0,k) +\Omega^{(\rm ind)}_{\rm gw}(\tau_0,k)\right) h^2\\
&\simeq \frac{\Omega_{r,0}\,h^2}{24}\left\{\mathcal{P}^{(s, {\rm p})}_{-} (\tau_i,k) + \sum_\lambda \left(\mathcal{P}^{(v, {\rm p})}_{\lambda} (\tau_i,k)+\left(\fr{k}{ \mathcal{H}(\tau_f)}\right)^2 \overline{\mathcal{P}^{(\rm ind)}_\lambda(\tau_f,k)}\right)\right\},
\end{align}where $\Omega_{\mathrm{r}, 0} h^2 \simeq 2.4 \times 10^{-5}$ is the radiation density today, $\tau_i$ represents a time right after inflation and $\tau_f$ denotes a time during radiation dominated universe such that $k\tau_f \gg 1$. For a detailed discussion on of each contribution that appear in \eqref{toto}, see Appendix \ref{AppD}. The quantity $\Omega_{\mathrm{gw}}(\tau_0,k) h^2$ is typically plotted with respect to the frequency $f = k/2\pi$, which is related to the number of e-folds during inflation by \cite{ Garc_a_Bellido_2016}
\beq\label{freq}
N =N_{p} -41.7 +\ln\left(\fr{k_{p}}{0.05~{\rm Mpc^{-1}}}\right)-\ln\left(\fr{f}{100~{\rm Hz}}\right)+\ln\left(\fr{H(N)}{H_{p}}\right),
\eeq
where $N_p$ corresponds to the e-folding number when the pivot scale left the horizon and the last term in \eqref{freq} takes into account the evolution of the Hubble rate during inflation.

\begin{figure}[t!]
\begin{center}
\includegraphics[scale=0.8]{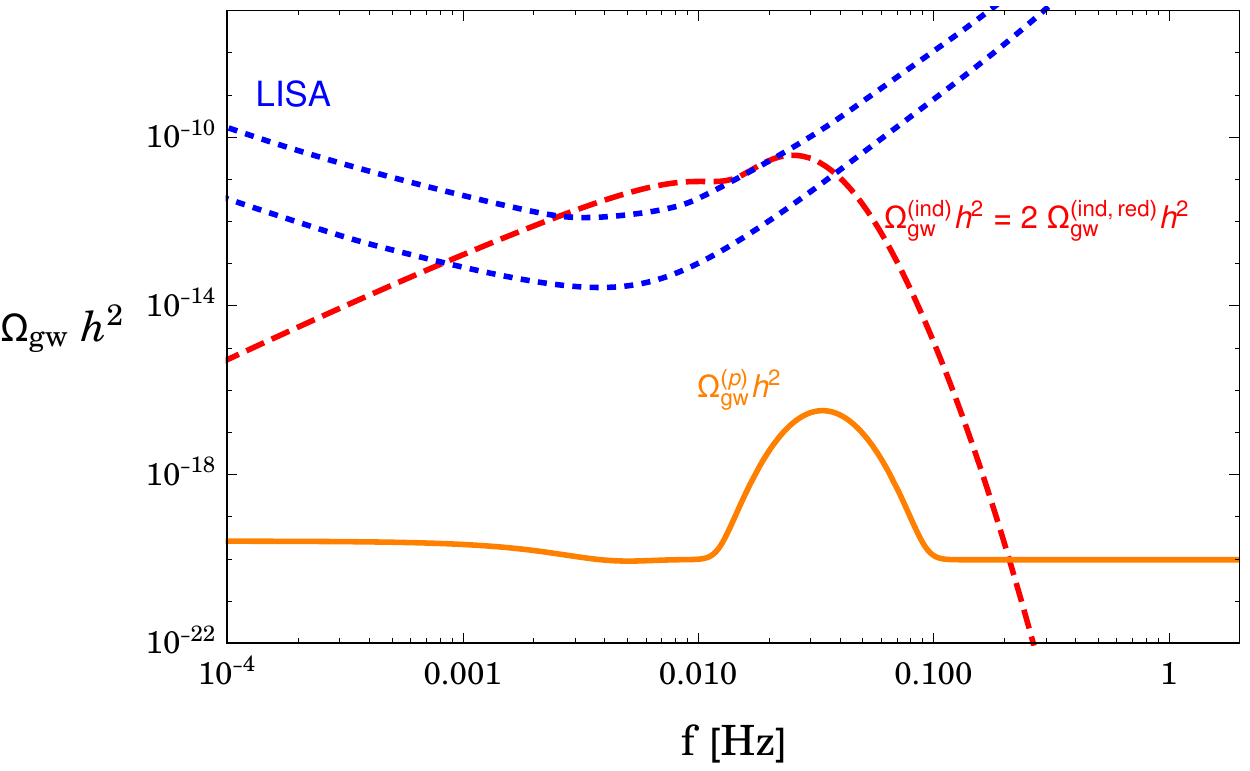}
\end{center}
\caption{Primordial (orange solid) and induced (red dashed) contributions to the total SGWB presented in eq. \eqref{toto} for the bumpy axion inflation. As explained in the main text, to estimate the total contribution to the induced signal, we multiplied the power spectrum of ``Reducible'' diagram $\mathcal{P}^{({\rm ind, red})}_{\lambda}$ by two (See \eg eq. \eqref{indps} of Appendix \ref{AppC}). \label{fig:omgcomp}}
\end{figure}

In the following, we compare primordial and induced part of the GW signal with the sensitivity curves of LISA\footnote{These sensitivity curves are shown in blue dotted lines in Figure \ref{fig:omgcomp}: A5M5 (bottom) and A2M2 (top) lines of Figure 1 of  \cite{Bartolo:2016ami}. In the notation AiMj, i refers to the length of the arms in millions of Km and j to the duration of the mission.} where we expect a peak in the spectrum to occur for the inflationary scenario we consider in Section \ref{Sec4p2}. In the present work, due to the non-Gaussian nature of scalar fluctuations sourced by the gauge fields, there are three distinct diagrams that contribute to the induced power spectrum \cite{Cai:2018dig,Unal:2018yaa}. To estimate the shape and amplitude of the resulting induced GW signal, we will only compute the dominant diagram we call ``Reducible'' and multiply this result by two in order to guess the final result (See the discussion in Appendix \ref{AppCC}.). We follow this route because the amplitude of the GW signal from the sum of other two diagrams (namely ``Planar'' and ``Non-Planar'') can at most be at the same order of magnitude compared to contribution arise from the ``Reducible'' diagram as shown previously in \cite{Garc_a_Bellido_2017}. 

In light of this information, we present both primordial and induced component that contributes to the total SGWB in Figure \ref{fig:omgcomp}. We observe that the primordial GW background $\Omega^{(\rm p)}_{\rm gw}$ that arise as a result of the parity breaking process $\delta A_- + \delta A_- \to h_-$ constitutes a completely sub-dominant portion of the total GW signal at LISA scales. The reason behind this is two folds: First and foremost, at much larger scales corresponding to $f \ll f_{\rm LISA} \simeq 10^{-3}\, {\rm Hz}$, amplitude of GWs are much smaller than the scalar fluctuations, in particular $ r \sim 10^{-5}$ at CMB scales corresponding to $\Omega_{\rm gw} h^2 \sim 10^{-19}- 10^{-20}$ for $f < f_{\rm LISA}$. This implies that approximately $10^{6}$ enhancement in the GW amplitude is required from direct sourcing of gauge fields for frequencies around $f \sim f_{\rm LISA}$. However, an amplification at this level is not allowed as the parameter $\xi_*$ that controls the particle production is bounded from above, which in turn restricts the maximum amplitude of sourced GWs can obtain. In particular, the theoretical bound on PBHs produced through this mechanism, namely the fact that PBH abundance should be less than the total dark matter abundance, \ie $F_{{\rm PBH}} \leq 1$ restricts the effective coupling to be $\xi_* \lesssim 10.54$ as we studied in Section \ref{Sec4p4p1}. 

{\bf Induced GWs at LISA scales:} On the other hand, we see from Figure \ref{fig:omgcomp} that the scalar fluctuations that are originally sourced by gauge fields during inflation can lead to a sizeable component of induced GWs visible at LISA scales. The double peak structure of the resulting GW spectrum, which is a typical behavior of induced GWs arising from scalar fluctuations exhibiting a narrow peak (such as a delta function), can be barely seen in Figure \ref{fig:omgcomp}. In the model we study here, the reason for this stems from the fact that scalar fluctuations exhibit a width that is slightly above the threshold value quoted in \cite{Pi:2020otn}, \ie $\sigma_{2,\mathcal{R}} \gtrsim \sigma_{\rm c} \sim 0.4$, to generate such a doubly peaked spectral shape as can be inferred from Table \ref{tab:fit1} using the limiting value of $\xi_* = 10.54$. It is worth emphasizing that, the same scalar fluctuations that generates the GW signal we study here can collapse into primordial black holes of mass $M \simeq 10^{-13} M_{\odot}$ (See Section \ref{Sec4p4p1}). Therefore, LISA measurements can shed light on such small PBHs and particularly to the inflationary mechanism that produces these objects.

Note that since PBH abundance is dictated by the ratio $\sqrt{\mathcal{P}_{\mathcal{R}}}/\mathcal{R}_c$ (See eq. \eqref{bpbh}), a decrease in $\mathcal{R}_c$ by a factor of $d$ would lead to the same PBH population if we reduce the scalar power spectrum by a factor of $d^2$. This in turn implies a $d^4$ decrease in the induced GW spectrum we present in Figure \ref{fig:omgcomp} as $\mathcal{P}^{(\rm ind)}_\lambda \propto \mathcal{P}_\mathcal{R}^2$. Comparing the maximum level of the induced GW signal (red dashed curve) with the lowest sensitivity curve of LISA in Figure \ref{fig:omgcomp}, we find that induced GW signal is below the sensitivity curve of LISA for $\mathcal{R}_c < 0.3$.

\subsection{Summary of results and comments}
\begin{itemize}[leftmargin=*]
\item In Section \ref{Sec4p2} and \ref{Sec4p3} we have seen that the presence of pronounced modulations in the axion potential (See eq. \eqref{Vb} and Figure \ref{fig:V}) alter inflationary dynamics in a way to provide sufficient amount of inflation even for an intermediate range of field excursions $\Delta \phi /\Mp\simeq \mathcal{O}(1)$ \cite{Parameswaran:2016qqq}. In particular, the existence of smooth plateaus in the potential leads to relatively small scale of inflation with a smaller tensor-to-scalar ratio $r \approx 10^{-5}$ at CMB scales when compared to models that exhibit smooth monomial terms in its scalar potential.

\item In Section \ref{Sec4p4p1}, we showed that in the presence of the coupling in eq. \eqref{LINT}, the motion of $\phi$ around the cliff-like region of its potential triggers an instability for vector fields which in turn efficiently amplify the curvature power spectrum through $\delta A_{-} + \delta A_{-} \to \mathcal{R}$, leading to a pronounced bump in the scalar power spectrum, see \eg right panel of Figure \ref{fig:ps}. We have seen that these scalar fluctuations can later collapse into PBHs of mass $M \simeq 10^{-13}\, M_{\odot}$ which can constitute the total dark matter abundance in the universe.

In Section \ref{Sec4p4p2}, we found that this large population of PBHs is accompanied by an unavoidable SGWB at LISA scales (See Figure \ref{fig:omgcomp}) due to the non-linear nature of gravity \cite{Mollerach:2003nq,Ananda:2006af,Osano:2006ew,Baumann:2007zm,Kohri:2018awv}. As a primordial mechanism that leads to these findings at sub-CMB scales, the strongly non-Gaussian nature of scalar fluctuations (which obeys $\chi^2$ statistics) in bumpy axion inflation can be considered as a distinguishing feature compared to single-field inflationary scenarios \cite{Franciolini:2018vbk,Passaglia:2018ixg,Atal:2018neu} and astrophysical backgrounds \cite{Bartolo:2018qqn} which are expected to be Gaussian to a high degree. For example, compared to a Gaussian model of peaked scalar fluctuations at sub-CMB scales, one requires a much smaller $\mathcal{P}_{\mathcal{R}}$ in the bumpy axion inflation to generate the same PBH abundance at the corresponding scales \cite{Garc_a_Bellido_2017}\footnote{In particular, scalar power spectra that generates the same PBH fraction $\beta$ are related through $\mathcal{P}_{\mathcal{R},\chi^2} \simeq 2 \mathcal{P}^2_{\mathcal{R},G}/\mathcal{R}_c^2$ where G stands for Gaussian origin of scalar fluctuations.}. Since the induced GW spectrum involve two copies of the enhanced scalar power spectrum, this in turn implies that the resulting induced GW spectrum will exhibit a smaller amplitude compared to an inflationary mechanism that generates a Gaussian bump in the scalar power spectrum. On the other hand, approximate double peak structure of the induced GW spectrum (See Figure \ref{fig:omgcomp}) we found in this work should be contrasted with the spectral shape of GWs generated in models that utilizes featureless monotic motion \cite{Domcke:2017fix,Barnaby:2011qe} and a transient relatively fast roll motion \cite{ Garc_a_Bellido_2016,Garc_a_Bellido_2017,Ozsoy:2020ccy} of axion-like fields during inflation. In contrast to the mechanism we studied in this work, in these models, the SGWB is dominated by the primordial component sourced directly by vector fields (\ie $\delta A + \delta A \to h$) where the GW spectrum exhibit a blue-tilted, monotonically increasing peak-less structure for a smooth featureless motion during axion inflation and a peaked log-normal shape for models that make use of transient fast-roll motion of a spectator axion-like field during inflation. In this context, signal reconstruction methods developed for the LISA mission \cite{Caprini:2019pxz} can be considered as a useful tool to distinguish the nature of inflationary mechanism that generates the GW signal. To sum up, the location, shape and amplitude of the induced GW spectrum together with the location and amplitude of the PBH mass distribution can provide experimental evidence on the inflationary mechanism responsible for this PBH population.
\item{{\bf Anisotropies of the SGWB:} Another observational consequence of the inflationary scenario we consider is anisotropies induced on the SGWB \cite{Bartolo:2019yeu}. In particular, in the present model, axion fluctuations can lead to position dependent effective coupling $\delta \xi$ which in turn can result with inhomogeneities of the primordial component of GW background \cite{Bartolo:2019oiq}. On the other hand, due to non-Gaussian nature of scalar perturbations in bumpy axion inflation we study here, a larger anisotropy might be produced for the induced GW component associated with PBH formation \cite{Bartolo:2019zvb}. Interestingly, both of these contributions to the GW anisotropy is controlled by the perturbation of $\xi$ which can be utilized to characterize the frequency dependence of the induced total anisotropy. We leave a detailed investigation on this matter for future work.}
\item{{\bf Implications on UV model building:} Finally, we would like to comment on the parameter space that leads to the sub-CMB phenomenology we discuss in this Section. We have seen that in the bumpy axion inflation model we study here, a large population of PBHs ($F_{\rm PBH} = 1$) and observable GWs of induced origin arise for an effective coupling $\xi_* \lesssim 10.54$ when the velocity of $\phi$ peaks during the rollover of the cliff-like region in its potential. Considering the relation $\xi_* = \alpha_{\rm c} \delta $ together with value of $\delta = 1.57$ implied by the background evolution we study in Section \eqref{Sec4p2}, the dimensionless coupling between the axion and gauge fields should take a value of $\alpha_{\rm c} \simeq 6.7$. We note that this value is smaller compared to the analysis appeared in \cite{Cheng:2016qzb,Cheng:2018yyr} where it was found that $\alpha_{\rm c} = 10-20$ is required to generate a significant population of PBHs and GWs at sub-CMB scales. Nevertheless, recent investigations suggest that a value of $\alpha_{\rm c} \simeq \mathcal{O}(1-10) $ could be hard to obtain in explicit string theory constructions on which axion monodromy models we are based on \cite{Barnaby:2011qe}. It would be interesting to identify explicit examples within type IIB string compactifications that give rise to $\alpha_{\rm c} \simeq \mathcal{O}(1-10)$. We leave investigations in this direction for a future work.}
\end{itemize}
\section{Conclusions and Outlook}\label{Sec5}

CMB and LSS observations provide strong evidence for primordial inflation. However, these observations allow us to access a small portion of the dynamics when compared with the total of $60$ e-folds required to solve the standard problems of Hot Big Bang cosmology. The remaining part of inflationary dynamics, corresponding to late times/smaller scales is yet to be fully explored apart from upper limits on the power of scalar fluctuations resulting from bounds on PBHs. PBHs can be considered as one of the possible experimental windows to probe inflationary physics at small scales. In light of current uncertainties of experimental bounds \cite{Kawasaki:2016pql,Niikura:2017zjd,Katz:2018zrn,Montero-Camacho:2019jte}, a possible mass window is around $M \simeq 10^{-13}\, M_{\odot}$ ($k \sim 10^{12} - 10^{13}\, {\rm Mpc^{-1}}$) for which PBHs could account for the total dark matter density in the universe. Interestingly, this mass window corresponds to modes produced around $N\sim 22$ before the end of inflation, corresponding to the optimal frequency $f \simeq \mathcal{O}({\rm 10^{-3} \,Hz})$ where LISA experiment will operate. 

In this work, we studied a string inspired mechanism of axion inflation that can generate a significant population of PBHs that can account for total DM abundance and observable GW signal of induced origin at scales/frequencies LISA mission is sensitive to. In particular, we showed that the motion of a non-compact axion-like field $\phi$ in its wiggly potential ($\Lambda^4 \lesssim m^2 f^2$) can experience transient fast roll(s) (with slow-roll violation) that can trigger a localized production of gauge field fluctuations that in turn generates an additional sourced component of enhanced scalar fluctuations required to produce PBHs at small scales (See Section \ref{Sec4p4p1}). Due to the ineludible coupling between tensor and scalar degrees of freedom at second-order in perturbation theory, the peaked scalar signal associated with PBH formation in this model also generate an observable SGWB at LISA scales whereas the primordial GW background directly sourced by gauge fields is sub-leading (See Section \ref{Sec4p4p2}).

We note that amplitude of the resulting induced GW signal can be considered as a direct probe of the statistics of the scalar perturbations produced during inflation \cite{Garc_a_Bellido_2017}: In the model we studied in this work, enhanced scalar perturbations originate from a convolution of two gauge field sources and hence obey $\chi^2$ statistics. This in turn imply that one requires a smaller amount of power in scalar fluctuations to produce the same amount of PBH population compared to an inflationary models that exhibit enhanced Gaussian scalar perturbations (See \eg models studied in \cite{Atal:2018neu}). Therefore, the resulting induced GW signal in the bumpy axion model we consider typically has a smaller amplitude compared to aforementioned models that exhibit nearly Gaussian scalar fluctuations. The spectral shape of the induced GW signal at LISA scales could also offer additional information on the origin of the mechanism that generates PBH dark matter: in the model we investigated, the shape of the GW signal near the peak region has a characteristic shape that stems from the marginally narrow peak structure of its scalar sources (See the discussion in Section \ref{Sec4p4p2}). The characteristic shape of induced GWs can thus serve as a distinguishing feature of the mechanism we study in this work, in particular compared to the inflationary scenarios aiming the produce observable GWs at small scales directly through spectator axion-gauge field dynamics \cite{ Garc_a_Bellido_2016,Garc_a_Bellido_2017,Ozsoy:2020ccy} and the models that exhibit a broad peak in the scalar perturbations \cite{Garcia-Bellido:2017mdw} which are expected to generate a smooth log-normal shape of induced GWs (See \eg \cite{Pi:2020otn}).

In the context of string-inspired model we are considering here, there remain to be several open questions. First of all, it would be interesting to initiate a scan of available parameter space that can lead to PBHs of mass $M \sim \mathcal{O}(10)\, M_{\odot}$ as for this mass range, the resulting GW signal is relevant at scales associated with future Pulsar Timing Array measurements \cite{Moore:2014lga} and hence can provide useful information \cite{ Garc_a_Bellido_2016}. On the other hand, in this work, we focused on scenarios where axion traverses a single bump during the entire inflationary expansion. Focusing on different parameter choices in bumpy axion inflation, it would be interesting the explore scenarios where multiple population of PBHs and observable GWs at sub-CMB scales can be generated.   A typical difficulty facing these scenarios is the fact that they must agree with CMB observations while keeping the interesting sub-CMB phenomenology intact. Another interesting venue that can be explored is to quantify the extent of gauge field production which could alleviate the fine tuning associated with PBH formation in single-field inflationary models \cite{Ballesteros:2017fsr,Hertzberg_2018}. In the context of string-inspired models, a good starting point for this analysis is to work with models that is capable of generating a large population of PBHs for which a significant tuning of potential parameters is required \cite{Ozsoy:2018flq}. We leave a comprehensive analysis on these issues for future work.
\acknowledgments
We would like to thank Caner \"Unal for illuminating discussions in the initial stages of this project. O\"O would also like to thank Guillermo Ballesteros, Susha Parameswaran, Gianmassimo Tasinato and Ivonne Zavala for useful conversations pertaining to this work. We are partially supported by National Science Centre, Poland OPUS project 2017/27/B/ST2/02531.

\begin{appendix}
\section{Background evolution and gauge field production through the bumps\label{AppA}}
\begin{figure}[t!]
\begin{center}
\includegraphics[scale=0.64]{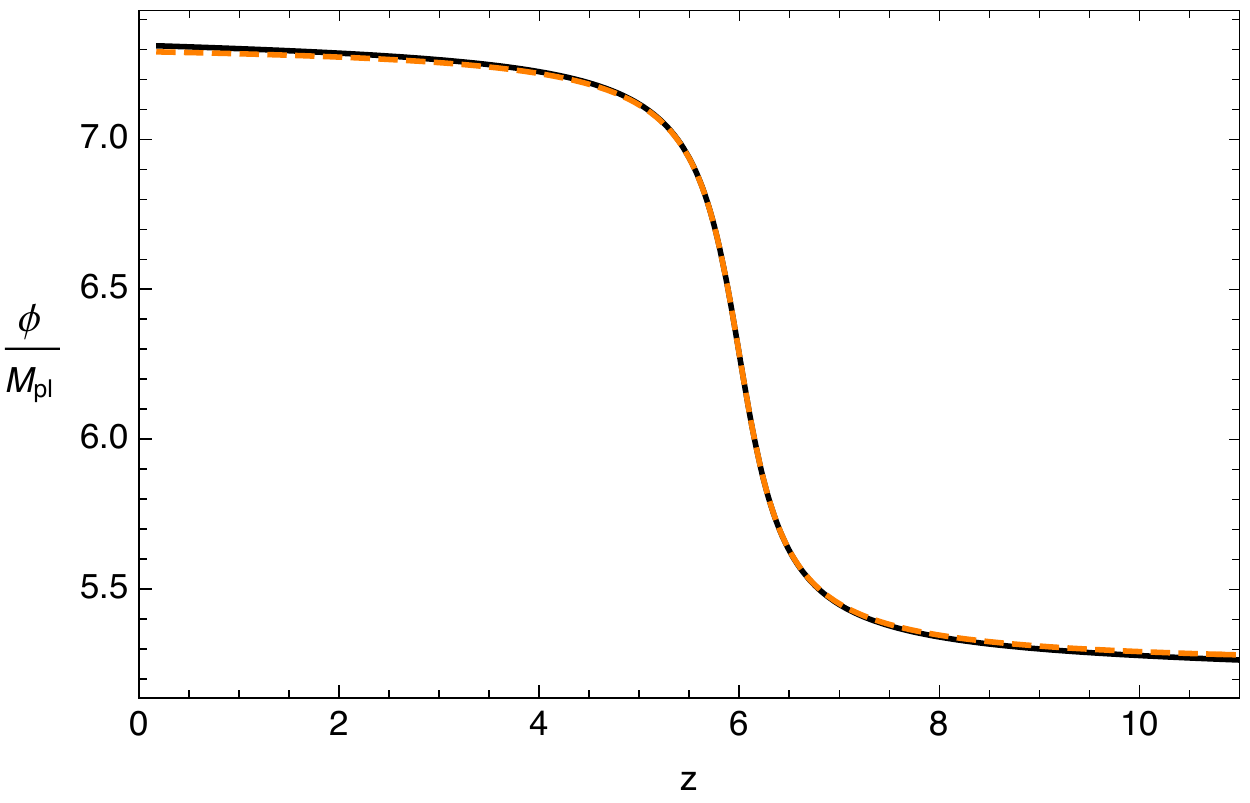}~\includegraphics[scale=0.64]{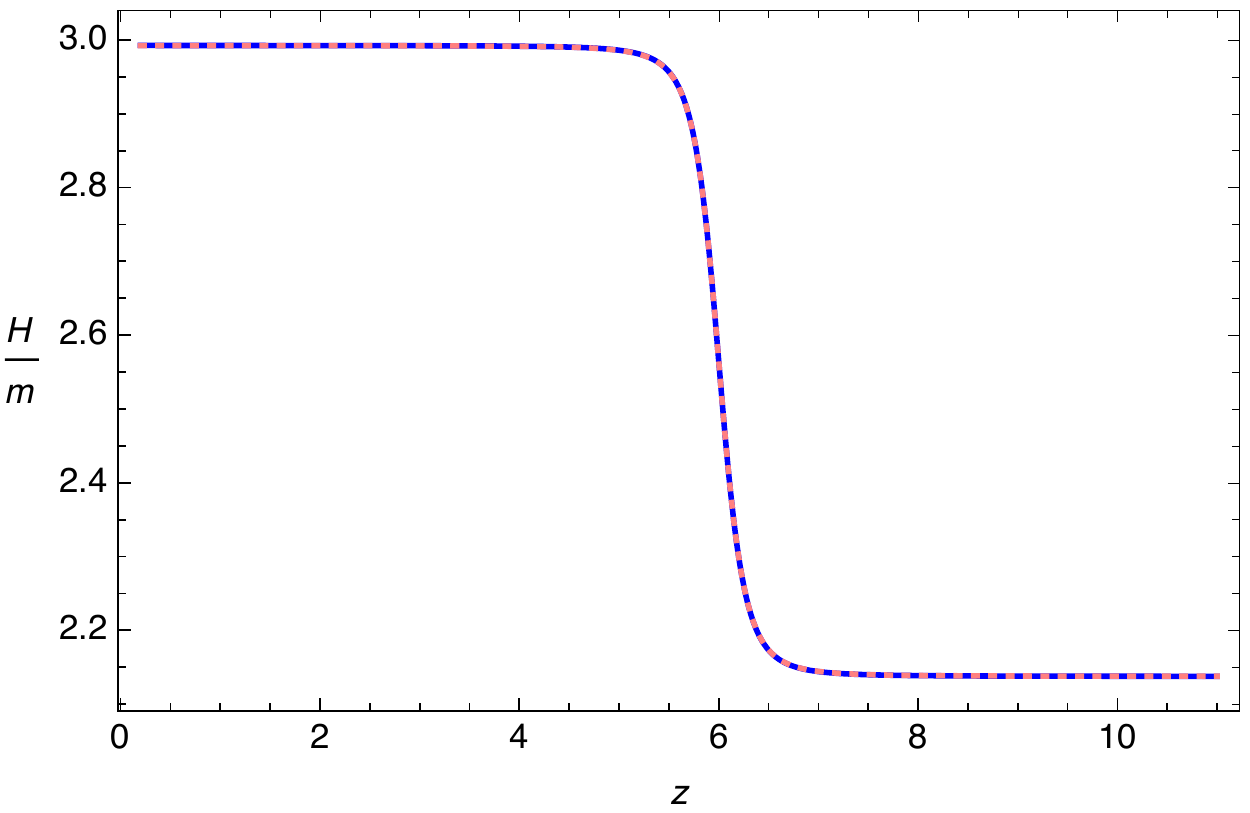}
\end{center}
\caption{The field profile $\phi$ \eqref{HJPS} and the Hubble parameter $H/m$ \eqref{HJH} as a function of $z = \sqrt{2/3} mt$ within a single bump of the potential \eqref{Vb}. In these plots, we take $\alpha =\Mp/f =  3$, $\beta=\Lambda^4 / (m^2 f^2) = 0.99$, $n = 6$ and $z_* =6$. In both panels, the resulting simplified profiles (dashed curves) are obtained using eq. \eqref{phip}.  \label{fig:HJb1}}
\end{figure}

\begin{figure}[t!]
\begin{center}
\includegraphics[scale=0.62]{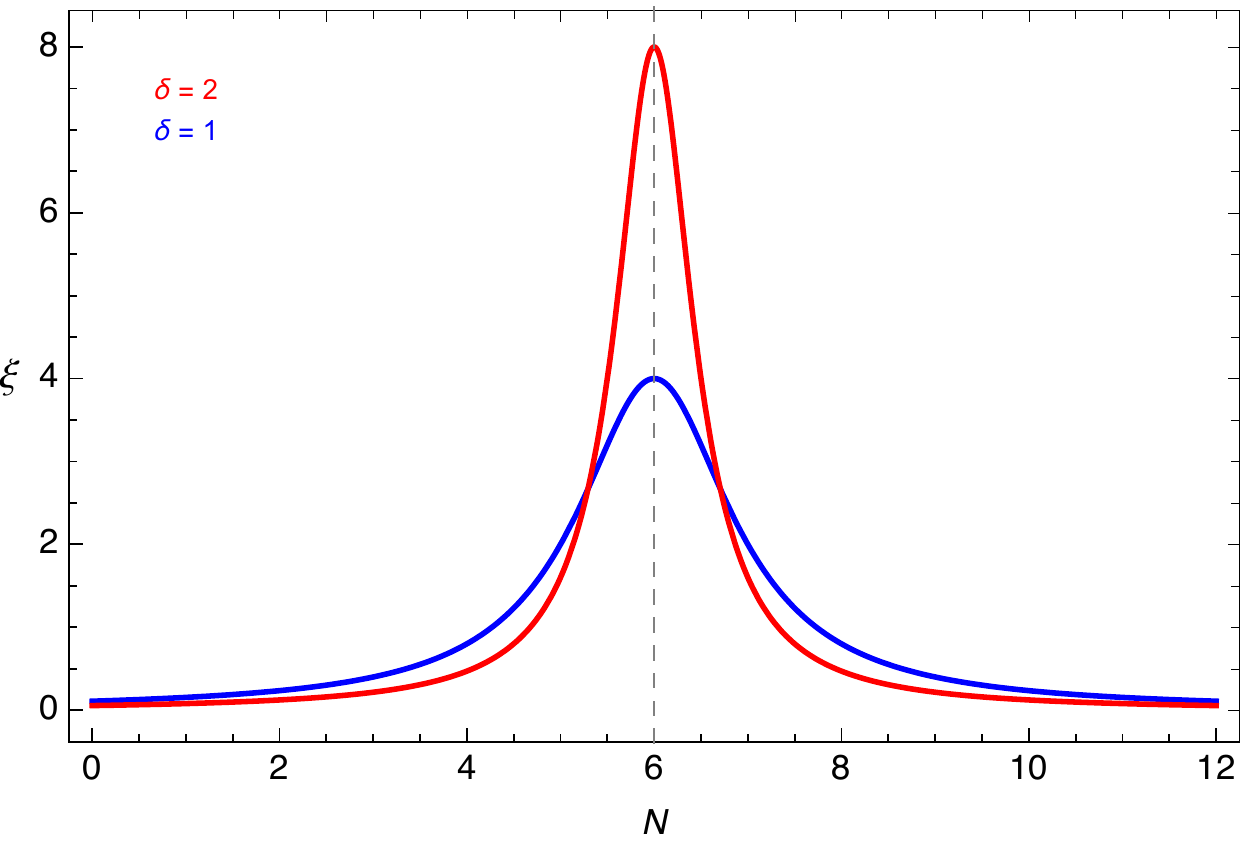}~\includegraphics[scale=0.65]{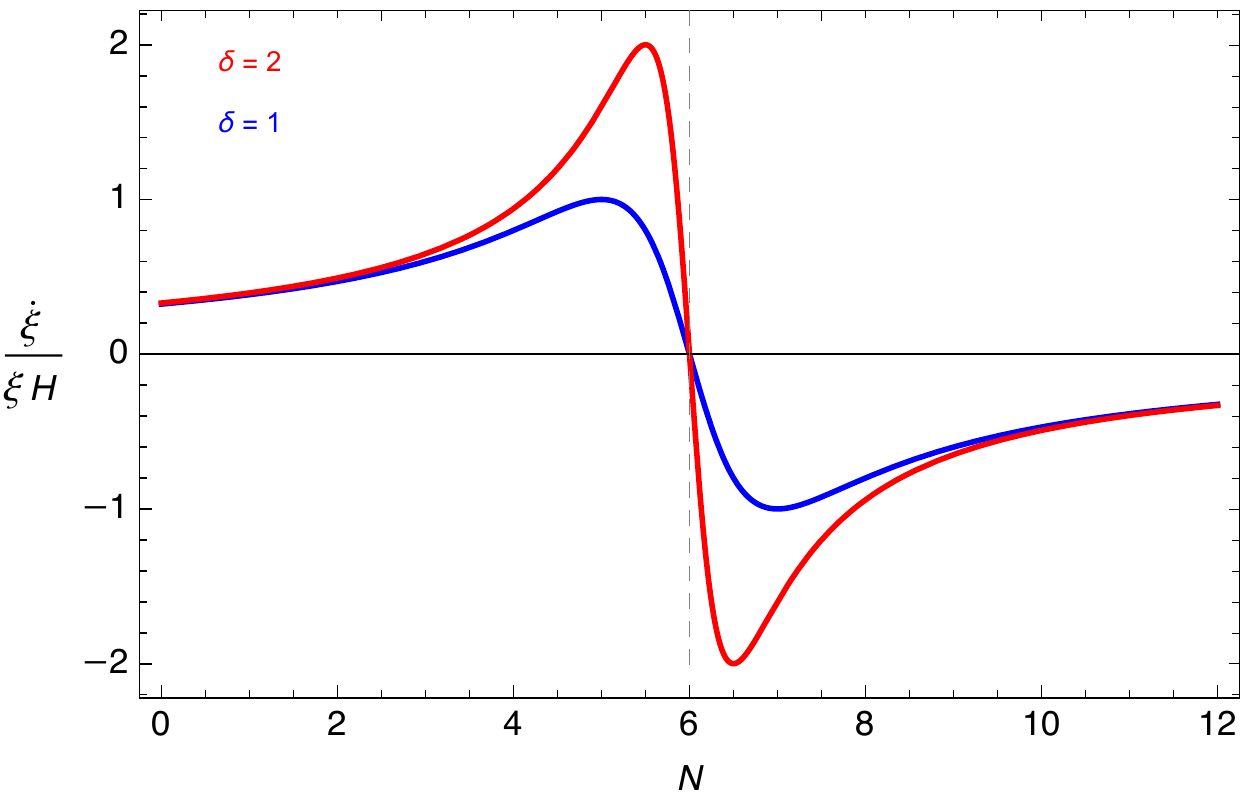}
\end{center}
\caption{The evolution of $\xi$ and $\dot{\xi}/\xi H$ as a function of e-folds within a bump of the potential \eqref{Vb}. In these plots, we take $\alpha_{\rm c} = 4$, $N_* = 6$.  \label{fig:xi}}
\end{figure}

In this appendix, our aim is to develop an analytic understanding of the scalar field profile as the inflaton rolls through gentle plateaus followed by steep cliffs. For this purpose, we use Hamilton-Jacobi approach \cite{Salopek:1990jq} (See also \cite{Parameswaran:2016qqq}) where the homogeneous background equations are given in terms of the scalar ``clock" field $\phi$ as
\begin{align}
\label{HJ1}-2 H'(\phi) ~\Mp^2 &= \dot{\phi}\\
\label{HJ2}3H^2(\phi)\Mp^2 &= 2H'^2(\phi) \Mp^4 + V(\phi),
\end{align}
where prime denotes differentiation with respect to the argument. Neglecting the kinetic energy of the scalar field, (which amounts to neglecting the first term on the right hand side of eq. \eqref{HJ2}, \ie $\epsilon < 1$), we can describe the Hubble parameter in terms of the scalar field $\phi$. At leading order in $\beta = \Lambda^4 / m^2 f^2 $, for sub-leading non-perturbative corrections in the potential \eqref{Vb}, we obtain\footnote{In particular, the leading order expression we derived in eq. \eqref{HJH} is valid for large enough scalar field values, \ie  $\alpha \tilde{\phi} \gg 1$ and for $\beta \lesssim 1$.}
\beq\label{HJH}
\fr{H(\tilde{\phi})}{m} \simeq \fr{\tilde{\phi}}{\sqrt{6}}\left[1+ \beta\,\, \fr{\sin(\alpha \tilde{\phi})}{\alpha \tilde{\phi}}\right] + \mathcal{O}(\beta^2),
\eeq
where we have defined the dimensionless field $\tilde{\phi}\equiv \phi/ \Mp$ and $\alpha \equiv \Mp /f$. 	Using the expression \eqref{HJH} for the Hubble rate in eq. \eqref{HJ1}, we derive a simple evolution equation for the scalar field,
\beq\label{HJP}
\tilde{\phi}'(z) + \left[ 1 + \beta \cos(\alpha \tilde{\phi}(z))\right] = 0,
\eeq
where we defined dimensionless time variable $z \equiv \sqrt{2/3} m t$. Notice that the equation \eqref{HJP} is invariant under the shift symmetry $\tilde{\alpha \phi} \to \tilde{\alpha \phi} + {2\pi n}$
for an arbitrary integer $n$. This implies that we can study the solution to \eqref{HJP} within the interval $(n-1){\pi} \leq \alpha\tilde{\phi} \leq (n+1)\pi$ 
for even $n$ and the remaining regions of the solution can be found using the periodicity of the eq. \eqref{HJP}. We thus make a field redefinition to study the evolution of the scalar field within such an interval, \ie for an even $n$, we write
\beq\label{HJPS}
\tilde{\phi}(z) = \fr{n\pi}{\alpha} + \fr{2}{\alpha} \arctan[y(z)],
\eeq
so that the new variable $y(z)$ obeys the following equation
\beq\label{eqy}
y'(z) + \fr{\alpha}{2} \bigg[1 + \beta + (1 - \beta) y^2(z)\bigg] = 0,
\eeq
The solution for $y$ is given by 
\beq\label{HJYS}
y(z) = \sqrt{\fr{1+ \beta}{1-\beta}}\tan \left[\fr{\alpha\sqrt{1-\beta^2}}{2}(z_*-z)\right],
\eeq
where $z_*$ is an integration constant. In the bumpy regime we are interested in, $\beta \to 1$, one can further simplify the solution in eq. \eqref{HJYS} as $y(z)\simeq {\alpha (1+ \beta)} (z_* - z)/2$
to obtain the scalar field profile in eq. \eqref{HJPS} as
\beq\label{phip}
\fr{\phi}{\Mp} = \fr{n\pi}{\alpha}  + \fr{2}{\alpha} \arctan\left[\fr{\alpha (1+\beta)}{2} (z_* - z)\right].
\eeq
In Figure \ref{fig:HJb1}, we present the accuracy of \eqref{phip} (shown by dashed lines) in describing the evolution of $\phi$ and Hubble rate $H$ \eqref{HJH} in comparison with the corresponding profiles obtained using \eqref{HJYS}. 
{\bf Gauge field production.} In the following, our aim is to derive approximate analytic formulas for the gauge field amplification when the inflaton rolls down through cliffs followed by plateau regions in its potential. From eq. \eqref{MEA}, we see that we need to determine an explicit expression for the time dependence of $\xi$ as $\phi$ traverses a single bump. For this purpose, we will neglect the time dependence of Hubble paramer $H$. For the model we are considering here, this simplifying assumption is justified by the fact that the gauge field production is mainly controlled velocity profile $\dot{\phi}$ in $\xi = -\alpha_{\rm c} \dot{\phi} / (2Hf)$ where the small change in $H$ around the cliffs does only affect the time dependence of $\xi$ marginally as can be verified from the right panel of Figure \ref{fig:HJb1} and from the field profile \eqref{phip} where $|\dot{\phi}|$ increases orders of magnitude. Keeping this in mind, we use \eqref{phip} and note $N = \ln a \simeq -\ln(-H\tau)$ to write $\xi$ as
\beq\label{xiapp}
\xi \equiv - \fr{\alpha_{\rm c}\,\dot{\phi}}{2Hf} = \fr{\alpha_{\rm c}~ \delta}{1+\ln\left[(x_*/x )^{\delta}\right]^2},
\eeq
where we defined the dimensionless ratio $\delta \equiv \alpha(1+\beta)(m/\sqrt{6} H)$ with $\alpha = \Mp/f$, $\beta = \Lambda^4/(m^2 f^2)$ and switched to $-k \tau = x$ where $\tau_*$ denoting the time at which $\xi$ reaches its peak value $\xi_* = \alpha_c\, \delta $. In Figure \ref{fig:xi},  we present time evolution of effective coupling $\xi$ and $\dot{\xi}/{\xi H}$ to show their sensitivity on the parameter $\delta$. We observe that larger $\delta$ results with a larger $\xi$ at fixed coupling $\alpha_{\rm c}$ at its peak, whereas its width reduces with increasing $\delta$. On the other hand, reducing $\delta$ significantly below unity, one can recover the adiabatic limit where $\dot{\xi}/\xi H \ll 1$.  In this work, our aim is to study gauge field production in the non-adiabatic regime for $\xi$, \ie for $\delta \simeq \mathcal{O}(1)$.
For this purpose, we use eq. \eqref{xiapp} in the mode equation \eqref{MEA} of the negative helicity mode to write
\beq\label{meam}
\fr{\d^2 A_-}{\d x^2} +\left(1 - \fr{2}{x}~ \fr{\xi_{*}}{1+\ln\left[(x_*/x )^{\delta}\right]^2} \right)A_{-} = 0.
\eeq
The late time growing solution to the eq. \eqref{meam} has been studied in detail in Appendix A of  which can be parametrized in terms of overall normalization factor as \cite{Ozsoy:2020ccy}:
\beq\label{solapp}
A_{-} \simeq N(\xi_*, x_*, \delta) \left[\frac{-\tau}{8 k \xi(\tau)}\right]^{1 / 4}  \exp \left[- \fr{2\sqrt{2\xi_*}~ (-k\tau)^{1/2} }{\delta |\ln(\tau/\tau_*)\,| }\right],\,\,\,\,\,\,\,\,\quad\quad\quad  \tau/\tau_* < 1,
\eeq
where the overall normalization $N(\xi_*, x_*, \delta)$ should be determined numerically which we compute by solving \eqref{meam} numerically and matching it to the WKB solution at late times $-k\tau \ll 1$. 

Focusing on values of $\xi_*$ within the range $8.5\leq \xi_*\leq 12$, we numerically solved \eqref{meam} for different values of $x_*$  and find that the normalization factor can be accurately described by the following shape
\beq\label{Nform}
N\left(\xi_{*}, q, \delta\right) \simeq N^{c}\left[\xi_{*}, \delta\right] \exp \left(-\frac{1}{2 \sigma^{2}\left[\xi_{*}, \delta\right]} \ln ^{2}\left(\frac{q}{q^{c}\left[\xi_{*}, \delta\right]}\right)\right),
\eeq
where the functions $N^{c}, q^c$ and $\sgm$ is characterized by the background evolution of $\phi$ which is parametrized by $\xi_*$ and $\delta$, \ie by its peak velocity and how fast the velocity reaches to its peak, respectively. We then match the late time amplitude obtained from the numerical solution of \eqref{meam} with the WKB solution in eq. \eqref{solapp}. In this way, we found that these functions can be described accurately by a second-order polynomial in $\xi_*$. In particular, for $\delta = 1.57$ we consider in this work, we obtained
\begin{align}
\nn N^{c} &=\exp \left(0.043+1.33 \,\xi_{*}-0.00073 \,\xi_{*}^{2}\right), \quad\quad \delta=1.57, \quad\quad 8.5 \leq \xi_{*} \leq 12, \\\nn
q^{c}&= 0.098 + 0.650 \,\xi_{*}-0.00033\, \xi_{*}^{2},\\
\sigma&=0.734 - 0.049\, \xi_{*} + 0.0014\, \xi_{*}^{2}.
\end{align}

\section{Primordial scalar power spectrum sourced by gauge fields}\label{AppC}
In this appendix, we present the derivation of the scalar power spectrum in the model \eqref{ML}. Using the fitting functions we devised for the gauge field mode functions, the results of this appendix can be used to obtain the phenomenology we discuss in Section \ref{Sec4}.

We start from \eqref{CP}, using the solution \eqref{Qphis} for the sourced canonical mode, the sourced curvature perturbation is given by
\beq
\hat{\mathcal{R}}^{(s)}(\tau,\vec{k}) =  \fr{H}{a\dot{\phi}}  \int^{\tau} d\tau'~ G^{\phi}_k(\tau,\tau')~ \hat{J}_\phi(\tau',\vec{k}),
\eeq
where the source is defined as in the right hand side of \eqref{Qphi}. Using the definitions \eqref{EBF}, it is given by
\begin{align}\label{Jps}
\nn\hat{J}_\phi(\tau',\vec{k}) & =\fr{\alpha_{\rm c}}{4f a(\tau')} \int \frac{\d^{3} p}{(2 \pi)^{3 / 2}} \,\epsilon^{-}_{i}(\vec{k}-\vec{p}) \epsilon^{-}_{i}(\vec{p})\,\, p^{1 / 4}\,|\vec{k}-\vec{p}|^{1 / 4}\left(p^{1/2}+ |\vec{k}-\vec{p}|^{1/2}\right) \\\nn\\
&\quad\quad\quad\quad\quad\quad\quad\quad\quad \times \tilde{A}(\tau',|\vec{k}-\vec{p}|)\, \tilde{A}(\tau', p)\, \hat{\mathcal{O}}_{-}(\vec{k}-\vec{p})\, \hat{\mathcal{O}}_{-}(\vec{p}),
\end{align}
where we symmetrized the integrand with respect to $p$ and $|\vec{k}-\vec{p}|$ and $\mathcal{O}_{-}$ is defined as $\hat{\mathcal{O}}_\lambda(\vec{q}) \equiv \left[\hat{a}_{\lambda}(\vec{q})+\hat{a}_{\lambda}^{\dagger}(-\vec{q})\right]$. 
In terms of homogeneous solutions of \eqref{Qphi} (See also \eqref{VMp}), $G_k^{\phi}$ is given by  
\beq\label{GFpApp}
G_{k}^{\phi}\left(\tau, \tau^{\prime}\right)=i \Theta\left(\tau-\tau^{\prime}\right)\left[Q^{(v)}_\phi(\tau, k)Q^{(v)^*}_\phi(\tau', k)-Q^{(v)^*}_\phi(\tau, k) Q^{(v)}_\phi(\tau', k)\right].
\eeq
As the scalar rolls down steep cliffs in its potential \eqref{Vb}, the effective mass term in \eqref{Qphi} is expected to deviate significantly from its slow-roll value, \ie $m_{\rm eff}^2 \simeq -2/\tau^2$, which in turn implies that we can no longer use vanilla slow-roll solutions for $Q^{(v)}_\phi(\tau, k)$ when we construct the Green's function in \eqref{GFpApp}.  Nevertheless, one can simplify the Green's function by factorizing the strongly scale dependent part. For this purpose, note that we would like to obtain the sourced curvature perturbation in the late time limit $-\tau \to 0$. We assume that the solutions $Q^{(v)}_\phi(\tau, k)$ to the homogeneous part of \eqref{Qphi} are real in this limit \footnote{This can be ensured by fixing the arbitrary initial phase of the mode functions $Q^{(v)}_\phi(\tau, k)$.}. In this case, the sourced solution in the late time limit $-k\tau \ll 1$ can be written as 
\beq\label{scp}
\hat{\mathcal{R}}^{(s)}(0,\vec{k}) = \fr{2 H}{a \dot{\phi}}Q^{(v)}_\phi(0,\vec{k}) \int_{-\infty}^{0} d\tau'~ {\rm Im}[Q^{(v)}_\phi(\tau',\vec{k})]~ \hat{J}_\phi(\tau',\vec{k}).
\eeq
Recalling \eqref{tA}, we plug the source in \eqref{Jps} to \eqref{scp} to obtain
\begin{align}\label{sR}
\nn \hat{\mathcal{R}}^{(s)}(0,\vec{k}) &= \fr{H Q^{(v)}_\phi(0,\vec{k})}{a \dot{\phi}}\fr{ H \alpha_{\rm c}}{2^{3/2} f k^{5/2}} \int \frac{\d^{3} p}{(2 \pi)^{3 / 2}} \,\epsilon^{-}_{i}(\vec{k}-\vec{p}) \epsilon^{-}_{i}(\vec{p})\,\, p^{1 / 4}\,|\vec{k}-\vec{p}|^{1 / 4}\left(p^{1/2}+ |\vec{k}-\vec{p}|^{1/2}\right) \\\nn
&\quad\quad\quad\quad\quad\quad\quad\quad\quad\times  N\bigg(\xi_*, -|\vec{k}-\vec{p}|\tau_*,\delta\bigg)N\bigg(\xi_*, -|\vec{p}|\tau_*,\delta\bigg)\, \hat{\mathcal{O}}_{-}(\vec{k}-\vec{p})\, \hat{\mathcal{O}}_{-}(\vec{p})\\
&\quad\quad\quad\quad\quad\quad\quad\quad\quad \times  \mathcal{I}_{\mathcal{R}}\bigg[\xi_*,x_*,\delta,\sqrt{\fr{|\vec{k}-\vec{p}|}{k}} +\sqrt{\fr{\vec{p}}{k}}\bigg],
\end{align}
where we defined the time integral of the source as 
\beq\label{Rs}
\mathcal{I}_{\mathcal{R}}\bigg[\xi_{*}, x_{*}, \delta, Q\bigg] \equiv \int_{0}^{\infty} \d x^{\prime}\, x' \,{\rm Im}[\tilde{Q}_{\phi}^{(v)}(x')]\, \exp \left[-\frac{2 \sqrt{2\xi_{*}}}{\delta}\frac{x'^{1/2}}{|\ln(x'/x_*)\,|} Q\right]
\eeq
where we again switched the dimensionless variables $-k\tau' = x'$ and also defined the dimensionless mode functions $\sqrt{2k}\, Q^{(v)}_\phi(\tau, k) \equiv \tilde{Q}^{(v)}_\phi(x)$. We now use \eqref{Rs} to compute the sourced scalar power spectrum. 	We define the total scalar power spectrum as
\beq\label{DRPS}
\frac{k^{3}}{2 \pi^{2}}\left\langle\hat{\mathcal{R}}(0, \vec{k}) \hat{\mathcal{R}}(0, \vec{k}^{\prime})\right\rangle \equiv \, \delta \left(\vec{k}+\vec{k}^{\prime}\right) \,  \mathcal{P}_{\mathcal{R}}(k). 
\eeq
Similar to the case with tensors, we separate the total scalar power spectrum as $  \mathcal{P}_{\mathcal{R}}(k) 
=  \mathcal{P}^{(v)}_{\mathcal{R}}(k) +  \mathcal{P}^{(s)}_{\mathcal{R}}(k)$ where
\beq\label{PS0}
\mathcal{P}^{(v)}_{\mathcal{R}}(k) = \lim_{\tau \to 0^{-}} \fr{k^3}{2\pi^2} \left(\fr{ H }{a\dot{\phi}}\right)^2 \big|Q^{(v)}_\phi(\tau,\vec{k})\big|^2 \equiv  \fr{k^3}{2\pi^2} \left(\fr{ H Q^{(v)}_\phi(0,\vec{k}) }{a\dot{\phi}}\right)^2,
\eeq
by our construction. Taking the 2-pt correlator of \eqref{sR} and using the Wick's theorem for the operators $\mathcal{O}_{-}$, the sourced power spectrum can be extracted from the definition \eqref{DRPS} as
\begin{align}\label{sPSs}
\nn \mathcal{P}^{(s)}_{\mathcal{R}}(k)  &= \mathcal{P}^{(v)}_{\mathcal{R}}(k)\fr{ H^2 \alpha_{\rm c}^2 }{16\pi^2 f^2} \int_{0}^{\infty} \d \tilde{p} \int_{-1}^{1} \d \eta \,\,\, \tilde{p}^{5/2}\,\,\,(1-2 \tilde{p} \eta+\tilde{p}^{2})^{1 / 4}  \left[ \tilde{p}^{1/2}+ (1-2 \tilde{p} \eta+\tilde{p}^{2})^{1 / 4}\right]^2 \\\nn
&\quad\quad\quad\quad\quad\quad\quad\quad\times  \big| \epsilon^{-}_{i}(\vec{k}-\vec{p}) \epsilon^{-}_{i}(\vec{p}) \big|^2N^2\bigg(\xi_*, (1-2 \tilde{p} \eta+\tilde{p}^{2})^{1 / 2}\,x_*,\delta\bigg)N^2\bigg(\xi_*, \tilde{p}\,x_*,\delta\bigg)\,\\
&\quad\quad\quad\quad\quad\quad\quad\quad\quad \times  \mathcal{I}^2_{\mathcal{R}}\bigg[\xi_*,x_*,\delta, (1-2 \tilde{p} \eta+\tilde{p}^{2})^{1 / 4}+\tilde{p}^{1/2}\bigg],
\end{align}
where we switched to dimensionless variable $\tilde{p} = p/k$ and $\eta$ denotes the cosine angle between $\vec{p}$ and $\vec{k}$. We express the overall normalization factor in \eqref{sPSs} in terms of the $\xi_*$ and $\epsilon_{\phi,*} = 2 \delta^2 /\alpha^2$ as
\beq\label{norm}
\fr{ H^2 \alpha_{\rm c}^2 }{16\pi^2 f^2} = \fr{H^2}{8\pi^2 \Mp^2} \,\fr{\xi_{*}^2}{\epsilon_{\phi,*}} 
\eeq
where we used $\xi(t) = ( \alpha_{\rm c} \Mp/f) \sqrt{\epsilon_\phi(t)/2}$ noting $\Mp/f \equiv \alpha$. Finally noting the following identity of polarization vectors, $\big| \epsilon^{\lambda}_{i}(\vec{p}) \epsilon^{\lambda'}_{i}(\vec{q}) \big|^2 = (1 -\lambda\lambda' \hat{p}.\hat{q})^2/4$, we write the total power spectrum as 
\beq
\mathcal{P}_{\mathcal{R}}(k) = \mathcal{P}^{(v)}_{\mathcal{R}}(k) \bigg[1 + \fr{H^2}{64\pi^2 \Mp^2} f_{2,\mathcal{R}}(\xi_*,x_*,\delta)\bigg],
\eeq
where we factorized all the effects containing gauge field production in the following function:
\begin{align}\label{f2Rf}
\nn f_{2,\mathcal{R}}(\xi_*,x_*,\delta) &= \fr{2\xi_{*}^2}{\epsilon_{\phi,*}} \int_{0}^{\infty} \d \tilde{p} \int_{-1}^{1} \d \eta \,\,\, \tilde{p}^{5/2}\,\,\,(1-2 \tilde{p} \eta+\tilde{p}^{2})^{1 / 4}  \left[ \tilde{p}^{1/2}+ (1-2 \tilde{p} \eta+\tilde{p}^{2})^{1 / 4}\right]^2 \\\nn
&\quad\quad\quad\quad\times \left[1+ \fr{\tilde{p}-\eta}{(1-2 \tilde{p} \eta+\tilde{p}^{2})^{1 / 2}}\right]^2N^2\bigg(\xi_*, (1-2 \tilde{p} \eta+\tilde{p}^{2})^{1 / 2}\,x_*,\delta\bigg)N^2\bigg(\xi_*, \tilde{p}\,x_*,\delta\bigg)\,\\
&\quad\quad\quad\quad \times  \mathcal{I}^2_{\mathcal{R}}\bigg[\xi_*,x_*,\delta, (1-2 \tilde{p} \eta+\tilde{p}^{2})^{1 / 4}+\tilde{p}^{1/2}\bigg].
\end{align}
Alternatively, we can switch to the variables $x = \tilde{p} + |\vec{k}-\vec{p}|/{k}$, $y = \tilde{p} - |\vec{k}-\vec{p}|/{k}$. In this case, we have
\begin{align}\label{f2RF}
 f_{2,\mathcal{R}}(\xi_*,x_*,\delta) &= \fr{\xi_{*}^2}{2\epsilon_{\phi,*}} \int_{1}^{\infty} \d x \int_{0}^{1} \d y \,\,\, \fr{(\sqrt{x+y}+\sqrt{x-y})^2\,(1-x^2)^2}{\sqrt{x+y}\sqrt{x-y}}\\\nn
&\quad\quad\quad\quad\times N^2\bigg(\xi_*, \fr{x-y}{2}\,x_*,\delta\bigg)N^2\bigg(\xi_*, \fr{x+y}{2}\,x_*,\delta\bigg) \mathcal{I}^2_{\mathcal{R}}\bigg[\xi_*,x_*,\delta, \fr{\sqrt{x-y}+\sqrt{x+y}}{\sqrt{2}}\bigg].
\end{align}
Similar to the case with tensor fluctuations, armed with the normalization factors $N(\xi,x_*,\delta)$ of gauge field mode functions, we can integrate $f_{2,\mathcal{R}}$ numerically. Final ingredient we need to achieve this is the behaviour of  $\tilde{Q}^{(v)}_\phi(x) = \sqrt{2k}\, Q^{(v)}_\phi$ that appear inside the integral we defined in \eqref{Rs}. This is what we turn next.

\smallskip
{\bf Solution for the canonical mode functions $\tilde{Q}^{(v)}_\phi(x)$.} In the inflationary background we consider in Section \ref{Sec4p2}, the dynamics proceeds through three successive phases including an initial slow-roll stage, followed by a short transient non-slow roll stage where $\dot{\epsilon}/{\epsilon H} \equiv \eta < 0$ which finally connects to a final slow-roll era before inflation terminates. The behaviour of canonical scalar field fluctuation in such a background is typically non-trivial and may lead to scale dependent behavior. In order to capture the full behavior of mode functions and hence the vacuum power spectrum of curvature perturbation in \eqref{PS0}, we will rely on the numerical methods we mentioned in Section \ref{Sec4p4p1}. On the other hand, to compute the sourced contribution \eqref{f2RF} to the scalar power spectrum, we will require the late time behavior of $\tilde{Q}^{(v)}_\phi(x)$ inside the integral \eqref{Rs}. This is because for $x' > 1$, the gauge field mode functions are highly suppressed as it is clear from the exponential factor appearing in \eqref{Rs}. In other words, the dominant contribution to the time integral in \eqref{Rs} stems from the $x' \to 0$ region of its integrand. Therefore, for all practical purposes, it is sufficient to determine the canonical mode functions in the $\tau \gg \tau_*$ region, namely well after $\dot{\phi}$ reaches its peak value corresponding to the final attractor slow-roll era. As in the model we discuss in Section \ref{Sec4p2}, we will model the final phase with a constant $\eta$ where the mode equation for $\tilde{Q}^{(v)}_\phi(x)$ takes the standard form:
\beq\label{QME}
\partial_x^2\tilde{Q}^{(v)}_\phi(x)+\left(1 - \fr{\nu^2 - 1/4}{x^2}\right)\tilde{Q}^{(v)}_\phi(x) = 0,\quad\quad\quad\quad x \ll x_*
\eeq
where $\nu^2 = (3+\eta)^2/4 \simeq constant$  \footnote{For backgrounds where the system spends an appreciable amount of time in the intermediate non-slow roll phase ($\eta < 0$), the late time solution we obtain in this section can be also extended to the phase of transient $\eta < 0$ phase thanks to the duality between the final slow-roll and the intermediate non-slow roll era \cite{Cai:2017bxr,Atal:2018neu,Ozsoy:2019lyy}. This duality clearly manifest itself in equation \eqref{QME}, noticing that the index $\nu^2$ is invariant under $\eta \to -6-\eta$.}. In this case, the equation \eqref{QME} has the well known solutions that reduces to the Bunch Davies vacuum in the $-k\tau \equiv x \gg 1$ limit,
\beq
\tilde{Q}^{(v)}_\phi(x) = i \sqrt{\fr{\pi x}{2}} H^{(1)}_\nu(x), 
\eeq
where we picked the arbitrary initial phase to ensure the decaying solution is imaginary in the late time $x \to 0$ limit as we advertised earlier. To evaluate \eqref{f2RF}, we will therefore explicitly use 
\beq\label{RsF}
\mathcal{I}_{\mathcal{R}}\bigg[\xi_{*}, x_{*}, \delta, Q\bigg] = \sqrt{\fr{\pi}{2}}\int_{0}^{\infty} \d x^{\prime}\, x'^{3/2} \,J_{\nu}(x')\, \exp \left[-\frac{2 \sqrt{2\xi_{*}}}{\delta}\frac{x'^{1/2}}{|\ln(x'/x_*)\,|} Q\right].
\eeq
For the model we consider in this paper, shortly after the end of non-slow roll era with $\eta < -6$, the $\eta$ parameter becomes constant, settling to $\eta = 0.3$ in the final slow-roll attractor phase (See \eg Figure \ref{fig:epsandeta}). For the calculation of the sourced power spectrum in this model we will therefore use $\nu = (3+0.3)/2$ in \eqref{RsF}.
\section{Induced tensor power spectrum during radiation dominated era}\label{AppCC}
The induced GWB is produced in the radiation dominated era upon horizon re-entry of the scalar fluctuations that were sourced by the gauge fields during inflation. In this appendix, we provide a detailed derivation of the tensor power spectrum that arise in the presence of enhanced scalar fluctuations in bumpy axion inflation. In terms of the canonical variable $Q_{\lambda}$ we defined in Section \ref{SF}, the relevant part of the action that accounts for this contribution is given by
\beq\label{SQInd}
S\left[\hat{Q}^{(\rm ind)}_{\lambda}\right]=\frac{1}{2} \int \d \tau \d^{3} k\left\{\hat{Q}^{(\rm ind)\,'}_{\lambda}\hat{Q}^{(\rm ind)\,'}_{\lambda}-\left[k^{2}- \fr{a''(\tau)}{a(\tau)}\right] \hat{Q}^{(\rm ind)^2}_{\lambda} + 2\hat{Q}^{(\rm ind)}_\lambda ~ \hat{J}^{(\rm ind)}_\lambda (\tau, \vec{k})\right\},
\eeq
which leads to the following equation of motion for the canonical variable,
\beq\label{indme}
\left(\partial^2_\tau + k^2 -\fr{a''(\tau)}{a(\tau)}\right)\hat{Q}^{(\rm ind)}_\lambda(\tau,\vec{k}) = \hat{J}^{(\rm ind)}_{\lambda}(\tau,\vec{k}),
\eeq
where $a(\tau) \propto \tau$ during radiation dominated universe (RDU) and the source is given by 
\cite{Ananda:2006af,Osano:2006ew,Baumann:2007zm,Kohri:2018awv}
\beq
\hat{J}^{(\rm ind)}_{\lambda}(\tau, \vec{k})=2\Mp a(\tau) \int \frac{\mathrm{d}^{3} p}{(2 \pi)^{3 / 2}}\Pi_{\lambda}(\vec{k},\vec{p})\, f(p\tau,|\vec{k}-\vec{p}|\tau)\, \hat{\mathcal{R}}(0,\vec{k}) \hat{\mathcal{R}}(0,\vec{k}-\vec{p}),
\eeq
where we defined $\Pi_{\lambda} (\vec{k},\vec{q})\equiv \Pi_{ij,\lambda}(\vec{k})\,q_iq_j$ and 
\beq\label{sf}
f(z,z') \equiv \fr{4}{9} \left(2 \,T(z) T(z') + \tilde{T}(z)\tilde{T}(z')\right),
\eeq
with $\tilde{T}(z) \equiv T(z) + z\, \partial_z T(z) $ where $T$ is the transfer function of metric perturbation in Newtonian gauge: $\Phi(\tau,\vec{k}) = (2/3) T(k\tau) \mathcal{R}(\tau,\vec{k})$ and is defined by
\beq
T(x)=\frac{9}{x^{2}}\left[\frac{\sin (x / \sqrt{3})}{x / \sqrt{3}}-\cos (x / \sqrt{3})\right].
\eeq
The sourced solution to the canonical variable is given by 
\beq
\hat{Q}^{(\rm ind)}_\lambda(\tau,\vec{k}) = \int^{\tau} \d \tau' \, G_k(\tau,\tau')\,  \hat{J}^{(\rm ind)}_{\lambda}(\tau',\vec{k}),
\eeq
where $G_k (\tau,\tau')$ is the Green's function of the homogeneous part of eq \eqref{indme} and is given by 
\beq
k G_k(\tau,\tau') = \sin (k(\tau-\tau')).
\eeq
\begin{figure}[t!]
\begin{center}
\includegraphics[scale=0.4]{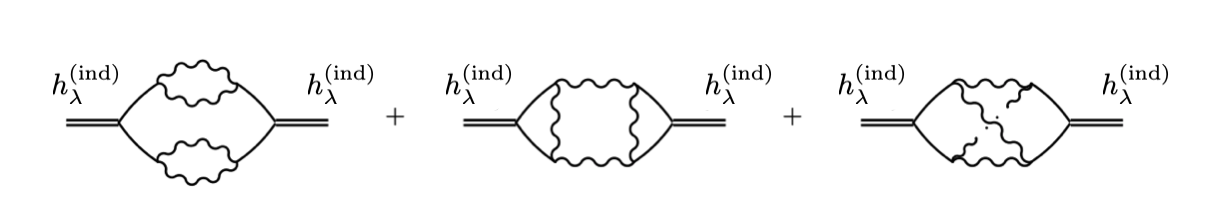}
\end{center}
\caption{Diagrams that contribute to the induced power spectrum of GWs in the bumpy axion monodromy inflation. Intermediate wiggly/solid lines represent vector field $A_-$ and scalar $\mathcal{R}$ fluctuations respectively. \label{fig:Dind}}
\end{figure} 
Noting the relation \eqref{rqtoh}, 2-pt correlator of induced tensor perturbation is given by
\begin{align}\label{indtpsgen}
\frac{k^{3}}{2 \pi^{2}}\left\langle\hat{h}^{(\rm ind)}_{\lambda}(\tau, \vec{k}) \hat{h}^{(\rm ind)}_{\lambda^{\prime}}(\tau, \vec{k}^{\prime})\right\rangle &= \fr{16}{2\pi^2 k}\int \fr{\d^3 p\, \d^3 q}{(2\pi)^3}\, \Pi_{\lambda} (\vec{k},\vec{p})\,\Pi_{\lambda'} (\vec{k'},\vec{q})\,\langle\hat{\mathcal{R}}_{\vec{p}} \,\hat{\mathcal{R}}_{\vec{k}-\vec{p}}\,\hat{\mathcal{R}}_{\vec{q}}\, \hat{\mathcal{R}}_{\vec{k'}-\vec{q}}\,\rangle\\\nn
&~~~~~~~~~~\times\int_0^{x} dx' \int_0^{x} dx'' kG_k(\tau,\tau')\, kG_k'(\tau,\tau'')\, \fr{a(\tau') a(\tau'')}{a(\tau) a(\tau)}\\\nn
&~~~~~~~~~~~~~~~~\times f(p\tau',|\vec{k}-\vec{p}|\tau')\, f(q\tau,|\vec{k'}-\vec{q}|\tau''),
\end{align}
where we introduced a shorthand notation for the curvature perturbation at the reheating surface as $\mathcal{R}(0,\vec{q}) \equiv \mathcal{R}_{\vec{q}}$. For a gaussian $\mathcal{R}$, the connected part of the 4-pt expectation value that appear in \eqref{indtpsgen} can be written as a sum two identical terms each containing 2-pt products of $\mathcal{R}$: \ie $\langle\hat{\mathcal{R}}_{\vec{p}} \,\hat{\mathcal{R}}_{\vec{k}-\vec{p}}\,\hat{\mathcal{R}}_{\vec{q}}\, \hat{\mathcal{R}}_{\vec{k'}-\vec{q}}\,\rangle \equiv 2 \langle\hat{\mathcal{R}}_{\vec{p}}\hat{\mathcal{R}}_{\vec{q}}\,\,\rangle\langle\hat{\mathcal{R}}_{\vec{k}-\vec{p}} \hat{\mathcal{R}}_{\vec{k'}-\vec{q}}\rangle $. In this case, using \eqref{DRPS}, induced tensor power spectrum can be simply written as a convolution of two scalar power spectrum (See \eg eq. (14) of \cite{Kohri:2018awv}). In the bumpy axion model we are focusing, the dominant contribution to the curvature perturbation is given by the part of the curvature perturbation $\mathcal{R}^{(s)}$ in \eqref{sR} that is sourced by two copies of amplified gauge fields and therefore it is highly non-Gaussian. As a result, using \eqref{sR}, one may realize that there are many different diagrams that can contribute to the induced GW spectrum for a $\mathcal{R}$ that obeys non-Gaussian statistics. Using all the possible contractions of gauge field raising and lowering operators that emerge from \eqref{sR} in \eqref{indtpsgen}, the diagrams that contribute to the induced GW spectrum are shown in Figure \ref{fig:Dind}. We label the first diagram on the left as ``'Reducible'' as in this case the 4-pt $\langle\mathcal{R}^4\rangle$ in \eqref{indtpsgen} can be written as a product of two sourced scalar power spectra in \eqref{sPSs} and therefore equivalent to the standard 1-loop computation that arise for Gaussian $\mathcal{R}$ we described above. The other two diagrams can be denoted as ``Planar'' and ``Non-Planar'' and must be evaluated through a 3-loop calculation. In a model that exhibit similar features with the model we consider here, these loop calculations involving integrals over internal momenta are calculated by approximating the width of the amplified gauge field functions (See \eg \eqref{Nform}) by a dirac delta distribution and the resulting contributions to the GW spectrum form these diagrams are found to be around the same order of magnitude for ``Planar'' and an order of magnitude lower for ``Non-Planar'' case compared to the ``Reducible'' diagram \cite{Garc_a_Bellido_2017}. In light of this information, in order to capture the overall spectral shape of the resulting induced GW signal, we will only focus on the ``Reducible'' diagram and multiply this result by two to determine its final amplitude. Using \eqref{sPSs} explicitly, the reducible contribution to the 4-pt function that appear in \eqref{indtpsgen} can be identified as
\beq\label{4pf}
\langle\hat{\mathcal{R}}_{\vec{p}} \,\hat{\mathcal{R}}_{\vec{k}-\vec{p}}\,\hat{\mathcal{R}}_{\vec{q}}\, \hat{\mathcal{R}}_{\vec{k'}-\vec{q}}\,\rangle = 2\delta(\vec{k}+\vec{k'})\delta(\vec{p}+\vec{q}) \fr{2\pi^2}{p^3} \mathcal{P}^{(s)}_{\mathcal{R}}(p) \fr{2\pi^2}{|\vec{k}-\vec{p}|^3} \mathcal{P}^{(s)}_{\mathcal{R}}(|\vec{k}-\vec{p}|)+ \dots,
\eeq
where dots represent the terms related to the planar and Non-Planar diagrams. Finally plugging eq. \eqref{4pf} in \eqref{indtpsgen} and noting the identity $\int \d \phi \,\, \Pi_{\lambda} (\vec{k},p) \,\Pi_{\lambda'} (-\vec{k},-\vec{p}) = \fr{p^4}{4} \left(1-\eta^2\right)^2 2\pi \, \delta_{\lambda\lambda'}$
where $\eta \equiv \hat{k}\cdot \hat{p}$, we extract the induced tensor power spectrum of the reducible diagram from the definition \eqref{DTPS} as \cite{Kohri:2018awv},
\beq
\mathcal{P}^{({\rm ind,red})}_{\lambda}(\tau, k)=4 \int_{0}^{\infty} \mathrm{d} v \int_{|1-v|}^{1+v} \mathrm{d} u\left(\frac{4 v^{2}-\left(1+v^{2}-u^{2}\right)^{2}}{4 u v}\right)^{2} I_{\rm ind}^{2}(u, v, x)\, \mathcal{P}^{(s)}_{\mathcal{R}}(k u) \mathcal{P}^{(s)}_{\mathcal{R}}(k v),
\eeq
where we switched to variables $u=|\vec{k}-\vec{p}| / k$ and $v=p / k$ and defined the time integral of the scalar sources as
\beq
I_{\rm ind}(u, v, x)=\int_{0}^{x} \mathrm{d} \bar{x}\, \frac{a\left(\bar{\tau}\right)}{a(\tau)}\, k G_{k}\left(\tau, \bar{\tau}\right) f\left(u\bar{x}, v\bar{x}\right).
\eeq
In order to evaluate the integrals, it is convenient to define $ t = u + v -1$ and $s = u - v$ to re-write the time averaged tensor power spectrum as 
\beq\label{indps}
\overline{\mathcal{P}^{({\rm ind,red})}_{\lambda}(\tau, k)} = 2 \int_{0}^{\infty} \mathrm{d} t \int_{-1}^{1} \mathrm{d} s\left[\frac{t(2+t)\left(s^{2}-1\right)}{(1-s+t)(1+s+t)}\right]^{2} \overline{I_{\rm ind}^{2}(u, v, x)}\, \mathcal{P}^{(s)}_{\mathcal{R}}(k u) \mathcal{P}^{(s)}_{\mathcal{R}}(k v),
\eeq
where $u=(t+s+1) / 2$ and $v=(t-s+1) / 2$. As we are interested in the induced GW signal today, we take the late time limit $x \gg 1$ of the oscillation average time integral $\overline{I^2_{\rm ind}}$ in \eqref{indps} which is given by \cite{Kohri:2018awv}
\begin{align}\label{Iind}
\nn \overline{I_{\mathrm{ind}}^{2}(t, s, x \rightarrow \infty)}=& \frac{288\left(-5+s^{2}+t(2+t)\right)^{2}}{x^{2}(1-s+t)^{6}(1+s+t)^{6}}\left(\frac{\pi^{2}}{4}\left(-5+s^{2}+t(2+t)\right)^{2} \Theta(t-(\sqrt{3}-1))\right.\\ &\left.+\left(-(t-s+1)(t+s+1)+\frac{1}{2}\left(-5+s^{2}+t(2+t)\right) \log \left|\frac{-2+t(2+t)}{3-s^{2}}\right|\right)^{2}\right).
\end{align}
In Section \ref{Sec4p4p2}, using \eqref{Indomg}, \eqref{Iind} and \eqref{indps}, we investigate the GW density resulting from the induced contribution we discussed in this appendix for the background model we focus in Section \ref{Sec4p2}.
\section{Energy density of stochastic GW backgrounds}\label{AppD}
In this appendix, we derive an expression for the fractional energy density of gravitational waves $\Omega_{\rm gw}$ with respect to the critical energy density. Inside the horizon (i.e when scales that exit during inflation re-enters the horizon), total energy density $\rho_{\rm gw} (\tau) = \int \d \ln k \,\rho_{\rm gw} (\tau, k)$ of gravitational waves is given by
\beq
\rho_{\mathrm{gw}}=\frac{\Mp^2}{4 a^{2}}\,\langle\overline{\partial_k h_{i j} \partial_k h_{i j}}\rangle,
\eeq
where overline denotes oscillation average for modes inside the horizon. Using the Fourier decomposition
\beq
{h}_{i j}(\tau, \vec{x})= \int \frac{\mathrm{d}^{3} k}{(2 \pi)^{3 / 2}} \mathrm{e}^{i \vec{k} \cdot \vec{x}} \sum_{\lambda=\pm} \Pi_{i j, \lambda}^{*}(\vec{k})\, {h}_{\lambda}(\tau, \vec{k})
\eeq
and noting the definition of the power spectrum \eqref{DTPS}, the fractional energy density of GWs is given by
\beq
\Omega_{\rm gw}(\tau,k) \equiv \frac{1}{\rho_c} \fr{\d \rho_{gw}}{\d \ln k} = \fr{\rho_{\rm gw}(\tau,k)}{3 H^2 \Mp^2} = \fr{1}{24}\left(\fr{k}{a(\tau) H(\tau)}\right)^2 \sum_{\lambda}\overline{\mathcal{P}_\lambda(\tau,k)}.
\eeq
Assuming that modes re-enter the horizon at radiation dominated universe (RDU), the GW energy density decays as radiation and so we can estimate the current energy density in terms of energy density in radiation today and $\Omega_{\rm gw}({\tau},k)$ where $\tau$ denotes a time during RDU\footnote{Note that in a radiation dominated universe, $a(\tau)H(\tau) \equiv \mathcal{H}(\tau) = \tau^{-1}$ with $\tau \geq 0.$} where the mode is deep inside the horizon $k\tau \gg 1$:
\beq\label{Omgfinal}
\Omega_{\rm gw}(\tau_0,k)\, h^2 = \frac{\Omega_{r,0}\,h^2}{24}\left(\fr{k}{ \mathcal{H}(\tau)}\right)^2 \sum_{\lambda}\overline{\mathcal{P}_\lambda(\tau,k)}.
\eeq
In the model we study in this work, there are various physical processes that contribute to the stochastic GW background (SGWB). Setting aside the primordial vacuum contribution $h^{(v,{\rm p})}$, we identify two distinct contributions to the metric perturbation that originates from vector field perturbations: {\emph i)}  The primordial component $h^{(s,{\rm p})} $ sourced directly by enhanced gauge fields during inflation and {\emph{ii})} the induced tensor perturbation $h^{(s,{\rm ind})}$ which originates from the enhanced scalar fluctuations (also sourced by vector fields during inflation) re-entering the horizon during RDU. Since the origin of these sources are different, we need to reinterpret the meaning of the formula \eqref{Omgfinal} suitably for each contribution. The primordial component of tensor fluctuations are generated during inflation via the process $\delta A_{-} + \delta A_{-} \to h_{-}$ and frozen at the reheating surface which then re-enters the horizon during RDU and evolves inside the horizon until today. Therefore, it is more suitable that we express this contribution in terms of its tensor power spectrum right after inflation ends as in this case modes are frozen. Noting that modes inside the horizon decay as $(k\tau)^{-2}$ for $k\tau > 1$  in \eqref{Omgfinal} \cite{Boyle:2005se}, the primordial component of SGWB density today is given by 
\beq\label{Pomg}
\nn \Omega^{(\rm p)}_{\rm gw} h^2 =  \frac{\Omega_{r,0}\,h^2}{24}  \sum_{\lambda} \left( \mathcal{P}^{(v, {\rm p})}_{\lambda} (\tau_i,k) + \mathcal{P}^{(s, {\rm p})}_{\lambda} (\tau_i,k)  \right)
\eeq
where we have removed the time average on the power spectrum as the primordial contribution $\mathcal{P}_\lambda (\tau_i, k)$ is of super-horizon origin with $\tau_i$ denoting an initial time in the RDU right after inflation ends. Note that for the model under consideration, both contributions in eq. \eqref{Pomg} are provided in Section \ref{ST}, see for example eq. \eqref{ptps}.

On the other hand, the induced component of SGWB arise as a result of amplified scalar fluctuations re-entering the horizon during RDU (namely through $\delta A_{-} + \delta A_{-} + \delta A_{-} + \delta A_{-} \to \mathcal{R} + \mathcal{R} \to h_{\pm}$) and hence involves sub-horizon evolution of its sources, namely the curvature perturbation (See \eg \eqref{sf} and the discussion it follows). Therefore, for the calculation of this contribution to the SGWB background, it is more convenient to evaluate the expression in \eqref{Omgfinal} at a reference time $\tau = \tau_f$ during RDU while the modes of interest are deep within horizon, \ie $k\tau_f  \to \infty$:
\beq\label{Indomg}
\Omega^{(\rm ind)}_{\rm gw}(\tau_0,k)\, h^2 = \frac{\Omega_{r,0}\,h^2}{24}\left(\fr{k}{ \mathcal{H}(\tau_f)}\right)^2 \sum_{\lambda}\overline{\mathcal{P}^{(\rm ind)}_\lambda(\tau_f,k)},
\eeq
where we kept the time average over the induced power spectrum to account for the oscillations of the scalar sources inside horizon. Combining the each contribution in eqs. \eqref{Pomg} and \eqref{Indomg}, the total fractional density of SGWB\footnote{The expression in eq. \eqref{totomg} neglects the contribution from the cross correlation of $h^{(s,{\rm p})}_{\lambda}$ and $h^{(s,{\rm ind})}_{\lambda}$. As we emphasized in Section \ref{Sec4p4p2}, the primordial contribution is already sub dominant in the axion inflation model we consider in this work and hence we neglect such cross terms that might appear in \eqref{totomg}. For a spectator axion model that can generate significant $h^{(s,{\rm p})}_{\lambda}$ and hence sizeable mixed correlators between $h^{(s,{\rm p})}_{\lambda}$ and $h^{(s,{\rm ind})}_{\lambda}$, see \eg \cite{Garc_a_Bellido_2017,Ozsoy:2020ccy}.} in the bumpy axion monodromy model is given by
\begin{align}\label{totomg}
\nn \Omega^{(\rm tot)}_{\rm gw}(\tau_0,k)\, h^2 &= \left( \Omega^{(\rm p)}_{\rm gw}(\tau_0,k) +\Omega^{(\rm ind)}_{\rm gw}(\tau_0,k)\right) h^2\\
&\simeq \frac{\Omega_{r,0}\,h^2}{24}\left(\mathcal{P}^{(s, {\rm p})}_{-} (\tau_i,k) + \sum_\lambda \mathcal{P}^{(v, {\rm p})}_{\lambda} (\tau_i,k)+\left(\fr{k}{ \mathcal{H}(\tau_f)}\right)^2 \sum_{\lambda}\overline{\mathcal{P}^{(\rm ind)}_\lambda(\tau_f,k)}\right),
\end{align}
where we have only taken into account the dominant helicity state $\lambda = -$ of the sourced primordial tensor perturbation that is sourced by $A_{-}$ and  the induced tensor power spectrum is given by eq. \eqref{indps} of Appendix \ref{AppCC}. 
\section{Backreaction analysis through the bumps}\label{AppE}
In this appendix, we will discuss the effects induced on the background motion of $\phi$ by the particle production in the gauge field sector. In particular, our aim is to find a valid parameter space in which the influence of vector field amplification on the inflation's motion can be neglected. In the mean field approximation, amplified the gauge fluctuations influence the evolution equation for the inflaton $\phi$ and the scale factor $\dot{a}/a = H$ through the following equations \cite{Barnaby:2011vw}, 
\begin{align}
\label{D1}\ddot{\phi} + 3 H \dot{\phi} &+ V'(\phi) = \fr{\alpha_{\rm c}}{f} \langle\vec{E}\cdot \vec{B}\rangle,\\
\label{D2}3 H^2 \Mp^2 &= \fr{1}{2} \dot{\phi}^2 + V(\phi) + \fr{1}{2} \langle \vec{E}^2 + \vec{B}^2 \rangle.
\end{align}
From \eqref{D1} and \eqref{D2}, to ensure that gauge fields have negligible effects on the background equations, we need to satisfy the following relations at any time during the background evolution,
\beq\label{brc}
\fr{1}{2}\langle\vec{E}^2+\vec{B^2}\rangle \equiv \rho_A \ll 3 H^2 \Mp^2,\quad\quad\quad\quad  3 H |\dot{\phi}| \gg \fr{\alpha_{\rm c}}{f} \langle\vec{E}\cdot \vec{B}\rangle.
\eeq
Notice that $|\vec{E}|/|\vec{B}| \simeq \sqrt{\xi/x} \sim \xi$ (see \eg \eqref{EBF}),  where we have used $x \sim \xi^{-1}$ for an optimal estimate on the latter ratio since for modes that satisfy $x \gg \xi^{-1}$, amplitude of mode functions is suppressed further (see eq. \eqref{tA}). Therefore, the second backreaction condition in \eqref{brc} can be re-written as
\beq\label{brc2}
\fr{\alpha_{\rm c} \langle \vec{E} \cdot \vec{B} \rangle}{f} \ll 3 H |\dot{\phi}| \quad\quad \longrightarrow\quad\quad \rho_ A \ll \, \fr{\dot{\phi}^2}{2},
\eeq
where we used the fact that $\vec{E}$ fields contribute dominantly to the energy density of the gauge fields as $|\vec{E}|/|\vec{B}| \simeq  \xi \simeq \mathcal{O}(10)$ to reach at interesting phenomenology in this work.  It is easy to realize that the condition appearing in \eqref{brc2} is more demanding compared to the first one appearing in \eqref{brc} and it simply guarantees that the energy density contained in the gauge field sector should be less than its reservoir, namely the kinetic energy of the inflaton. In the following, we will use \eqref{brc2} to derive the backreaction constraints on model parameters. 

Using the definition in eq. \eqref{brc} and expressions for electromagnetic fields in eq. \eqref{EBF}, the energy density in the gauge field sector can be parametrized as \cite{Namba:2015gja,Ozsoy:2020ccy}
\beq\label{rhoA}
\fr{\rho_{A}}{\epsilon_{\phi,p}\, \rho_\phi} = \fr{\mathcal{A}_s \,y^{7/2} N^{c}[\xi_*,\delta]^2 \sqrt{2\xi(y)}}{3}\int_{0}^{\infty} \d x_*\, x_*^{5/2} \exp\left[-\fr{4\sqrt{2\xi_*y}\,x_*^{1/2}}{\delta |\ln (y)|}-\fr{\ln(x_*/q_c)}{\sgm^2}\right]\left(1 + \fr{x_*\,y}{2\xi(y)}\right),
\eeq
where $y \equiv \tau/\tau_*$ and  $\mathcal{A}_s \equiv H^2/(8\pi^2 \epsilon_{\phi,p}\Mp^2) \simeq 2.1 \times 10^{-9}$ denoting the normalization of the power spectrum at CMB scales. Plugging \eqref{xiapp} into \eqref{rhoA} (and noting $\alpha_c \delta \equiv \xi_*$), $y = \tau/\tau_*$ dependence of the expression \eqref{rhoA} can be studied for different $\xi_*$ values. In this way, we found that at fixed $\xi_*$, the energy density in the gauge fields reaches a maximum around $y = \mathcal{O}(0.1)$ and quickly decays away both in the IR $\tau/\tau_* \to 0$ and UV $\tau/\tau_* \to \infty$ limits \cite{Namba:2015gja,Ozsoy:2020ccy}. On the other hand, for higher values of $\xi_*$, the maximum value reached by the expression in eq. \eqref{rhoA} increases due to the more efficient amplification of vector field modes for larger effective coupling $\xi_*$. 
At its maximum value, we studied $\xi_*$ dependence of $\rho_A / (\epsilon_{\phi,p}\, \rho_\phi)$ and found that it can be described accurately by the following expression,
\begin{align}\label{rhoAf}
\fr{\rho_{A,*}}{\epsilon_{\phi,p}\, \rho_\phi} &\simeq 1.25 \times 10^{-11} \,\,e^{2.533\, \xi_*},\quad\quad\quad \delta = 1.57.
\end{align}
Now realize that at the peak of the sourced signal, the back-reaction constraint \eqref{brc2} can be written as $\rho_{A,*} \ll \rho_\phi \epsilon_{\phi_*}/3$.
Using eq. \eqref{rhoAf}, this expression turns into
\begin{align}\label{rhoAf2}
1.25 \times 10^{-11} \,\,e^{2.533\, \xi_*} & \ll \fr{1}{3}\fr{\epsilon_{\phi,*}}{\epsilon_{\phi,p}},\quad\quad\quad \delta = 1.57.
\end{align}
In order to evaluate the right hand side of \eqref{rhoAf2}, we note $\epsilon_\phi = 2\delta^2/(\alpha (1 + \delta^2 \Delta N^2)^2$
where $\Delta N = N_p - N_*$. Finally as $N_p \simeq 55.6$ in the model we focus, we express eq. \eqref{rhoAf2} in terms of an upper bound on $\xi_*$ at scales where the gauge field production peaks, \ie around $N_* \simeq 24$:
\begin{align}
\xi_* < 15.6,\quad\quad\quad\quad\quad \delta = 1.57,
\end{align}
where we turned $\ll$ signs into $<$ due to exponential sensitivity to the parameter $\xi_*$. 
 \end{appendix}
 
\addcontentsline{toc}{section}{References}
\bibliographystyle{utphys}

\bibliography{paper2}

\end{document}